\documentclass[11pt,UKenglish]{article}
\usepackage[latin1]{inputenc}
\usepackage{babel}
\usepackage[a4paper,heightrounded]{geometry}
\usepackage{amsmath}
\usepackage{amsfonts}
\usepackage{amssymb}
\usepackage{bm}
\usepackage{graphicx}
\usepackage[numbers,sort&compress]{natbib}
\usepackage{varioref}
\usepackage{subfig}
\usepackage{tikz}
\usepackage[vcentermath]{myyoungtab}
\usepackage[bookmarksopen]{hyperref}

\allowdisplaybreaks % Umbrueche in Formeln erlauben
\Yinterspace{0.2ex plus 0.1ex} % Abstand neben Young-Tableaux

\DeclareMathOperator{\Tr}{Tr} % Spurbildung
\newcommand{\widebar}[1]{\overline{#1}} % Langer Oberstrich
\newcommand{\im}{\mathrm i} % Imaginaere Einheit
\newcommand{\ex}[1]{\mathrm e^{#1}} % Exponentialfunktion
\newcommand{\dif}{\mathrm d} % Ableitung
\newcommand{\opr}[1]{\mathcal{#1}} % Operatoren
\newcommand{\ket}[1]{|{#1}\rangle} % Ket-Vektor
\newcommand{\set}[1]{\mathbb{#1}} % Menge
\newcommand{\msp}{\,} % Abstand zu Satzzeichen in Formeln
\newcommand{\QTM}{\opr T^\text{QTM}}
\newcommand{\vecl}[1]{\mathbf{#1}} % Log-Vektor
\newcommand{\vect}[1]{\bm{#1}} % Vektor
\newcommand{\matr}[1]{\underline{\mathbf{#1}}} % Matrix
\newcommand{\vbethe}[1]{v_{k_{#1}}^{#1}} % Bethe-Ansatzzahl
\newcommand{\rep}[2]{{#1}^{({#2})}} % Funktion zu bestimmter Darstellung
\newcommand{\ord}{\mathcal O} % Ordnung
\newcommand{\alg}[1]{\mathit{#1}} % Algebra

\title{Nonlinear integral equations for the thermodynamics of the
  $\alg{sl}(4)$-symmetric Uimin-Sutherland model}
\author{Jens Damerau\thanks{e-mail:
    \href{mailto:damerau@physik.uni-wuppertal.de}{\protect\nolinkurl{damerau@physik.uni-wuppertal.de}}} \ and
  Andreas Kl\"umper\thanks{e-mail:
    \href{mailto:kluemper@physik.uni-wuppertal.de}{\protect\nolinkurl{kluemper@physik.uni-wuppertal.de}}}\\
  \parbox{0.9 \linewidth}{\vspace{0.4 \baselineskip}\centering
    Fachbereich C -- Physik, Bergische Universit\"at Wuppertal,\\
    42097 Wuppertal, Germany}}
\date{19th October 2006}

\begin{document}

\maketitle

\begin{abstract}
  We derive a finite set of nonlinear integral equations~(NLIE) for the
  thermodynamics of the one-dimensional $\alg{sl}(4)$-symmetric
  Uimin-Sutherland model. Our NLIE can be evaluated numerically for
  arbitrary finite temperature and chemical potentials. We recover the
  NLIE for $\alg{sl}(3)$ as a limiting case. In comparison to other recently
  derived NLIE, the evaluation at low temperature poses no problem in our
  formulation. The model shows a rich ground-state phase diagram. We obtain the
  critical fields from the $T \to 0$ limit of our NLIE. As an example for the
  application of the NLIE, we give numerical results for the $\alg{SU}(4)$
  spin-orbital model. The magnetic susceptibility shows divergences at
  critical fields in the low-temperature limit and logarithmic singularities
  for zero magnetic field.\\[\baselineskip]
\end{abstract}

\section{Introduction}

Since Bethe's seminal solution of the one-dimensional spin-1/2 Heisenberg
chain~\cite{bethe31}, many integrable, natural generalisations of this model
have been treated using basically the same ansatz. Among them is a
multi-component, higher-rank generalisation of the Heisenberg chain, first
proposed by Uimin for the case of three components~\cite{uimin70}. Later,
Sutherland introduced and solved the model for an arbitrary number of particle
types~\cite{sutherland75}. The two-dimensional classical model associated
with the one-dimensional Uimin-Sutherland~(US) model is the Perk-Schultz~(PS)
model~\cite{perk81}. The case of higher-rank representations of the underlying
symmetry algebra was treated by Andrei and
Johannesson~\cite{andrei84,johannesson86a}. Affleck calculated the critical
behaviour based on non-Abelian bosonisation and conformal field
theory~\cite{affleck86a,affleck88}.

The traditional thermodynamic Bethe ansatz~(TBA) allows for the treatment of
finite-temperature properties~\cite{yang69,yang70,takahashi71,gaudin71}. It
uses the string hypothesis and typically yields an infinite set of nonlinear
integral equations~(NLIE). The numerical solution of these equations poses a
problem as some kind of truncation scheme is necessary. Only in the limit $T
\to 0$ one obtains a finite set of equations. Using the TBA approach for the
general US model~\cite{johannesson86b}, the low-field asymptotics of the
susceptibility~\cite{schlottmann92} as well as the low-temperature asymptotics
of the specific heat~\cite{lee94a,lee94b} have been derived analytically.

With the help of the quantum transfer matrix~(QTM)
formalism~\cite{suzuki85,kluemper92b} and the fusion hierarchy of transfer
matrices~\cite{kirillov87,bazhanov90,kluemper92a,kuniba94,tsuboi97,tsuboi98}, it is
possible to rederive the TBA equations without using the string
hypothesis~\cite{juettner98}. Using the QTM it is also possible to derive an
alternative set of NLIE consisting only of a finite number of unknown
auxiliary functions~\cite{kluemper90,kluemper91,destri92,kluemper93}, thus
allowing for a precise numerical treatment at arbitrary finite
temperature. Later it was realised that these auxiliary functions provide a
natural way to exactly truncate the TBA
equations~\cite{suzuki99}. Nevertheless, no straightforward way of getting the
required auxiliary functions is known. Up to now, NLIE of this type have only
been derived for three components at
most~\cite{juettner97a,juettner97b,fujii99}.

There exists yet another type of NLIE~\cite{takahashi01}, which allows for the
generalisation to an arbitrary number of
components~\cite{tsuboi03}. Unfortunately, these NLIE prove difficult to
evaluate at low temperature. Instead, high-temperature expansions~(HTE) have
been obtained up to high order~\cite{shiroishi02,tsuboi03}. Very recently,
these equations have been further generalised to treat
$U_q(\alg{\widehat{sl}}(r|s))$-symmetric PS models~\cite{tsuboi06}.

In this paper, we treat the four-state, $\alg{sl}(4)$-symmetric US model in the
spirit of~\cite{kluemper93}. We define 14 suitable auxiliary functions, from which
we derive a set of well-posed NLIE that are valid for arbitrary finite
temperature and chemical potentials. The auxiliary functions are connected to
the fundamental representations of $\alg{sl}(4)$. Our NLIE are the natural
generalisation of those obtained for the $\alg{sl}(3)$-symmetric
case~\cite{fujii99}.

The $\alg{sl}(4)$-symmetric US model has many interesting applications. As an
example, we treat the $\alg{SU}(4)$ spin-orbital
model~\cite{yamashita98,yamashita00}. The thermodynamic properties of this
model have already been studied numerically using various
methods~\cite{frischmuth99,fukushima02,sirker04,tsuboi06}. The ground-state
phase diagram in dependence of the magnetic field and the orbital Land\'e
factor has also been obtained~\cite{gu02}.
In comparison to these methods, we are able to obtain highly accurate
numerical results for low finite temperatures in the thermodynamic
limit. Other possible applications of the $\alg{sl}(4)$-symmetric US model
include an integrable two-leg spin ladder system~\cite{wang99}, which has
recently been studied numerically using the HTE and TBA
methods~\cite{batchelor03a,batchelor03b}.

The outline of this paper is as follows. In Section~\ref{sec:usmodel}, we
briefly introduce the $q$-state US model and show how its thermodynamic
properties can be obtained using the QTM approach. In
Section~\ref{sec:nliesl4}, we concentrate on the $\alg{sl}(4)$-symmetric
case. We present a set of 14 well-posed auxiliary functions, from which we
derive a closed set of NLIE. We show how the largest eigenvalue of the QTM
can be extracted from these auxiliary functions. In
Section~\ref{sec:analytical}, we treat two limiting cases of our NLIE. First,
the NLIE and auxiliary functions of the $\alg{sl}(3)$-symmetric case are
recovered by freezing out one of the states. Second, the limit $T \to 0$
yields linearised integral equations, which are equivalent to the
corresponding TBA equations and allow for the derivation of the critical
fields. In Section~\ref{sec:numerical}, we deal with the numerical solution of
our NLIE. We briefly introduce the $\alg{SU}(4)$ spin-orbital model as an
application and give results for various physical quantities. In
Section~\ref{sec:summary}, we give a summary of our work and an outlook on
open problems. Appendix~\ref{app:nliederiv} is devoted to details concerning
the derivation of the NLIE.

\section{QTM approach to the US model}\label{sec:usmodel}

In order to fix notation, we begin with a short review of the
Uimin-Sutherland~(US) model~\cite{uimin70,sutherland75}. Consider a
one-dimensional lattice with $L$ sites, where a $q$-state spin variable
$\alpha_j$ is assigned to each site $j$. Each spin $\alpha$ has its own
grading $\epsilon_\alpha = (-1)^{p(\alpha)} = \pm 1$. The Hamiltonian of the
US model is then given by
\begin{equation}\label{eq:ushamilton}
  \opr H_0 = \sum_{j=1}^L \pi_{j,j+1}\msp,
\end{equation}
where the local interaction operator $\pi_{j,j+1}$ permutes neighbouring spins
on the lattice with respect to their grading,
\begin{equation}
  \pi_{j,j+1} \ket{\alpha_1 \ldots \alpha_j \alpha_{j+1} \ldots \alpha_L} =
  (-1)^{p(\alpha_j) p(\alpha_{j+1})} \ket{\alpha_1 \ldots \alpha_{j+1} \alpha_j \ldots \alpha_L}\msp,
\end{equation}
and periodic boundary conditions are imposed. The model shows
$\alg{sl}(r|s)$ symmetry, where $r$ and $s$ are the total number of states with positive and
negative grading ($q = r + s$), respectively. It is therefore a higher rank
generalisation of the spin-$1/2$ Heisenberg chain, which is contained as the
special case $q = 2$ and $\epsilon_1 = \epsilon_2 = +1$. 

\begin{figure}
  \centering
  $R_{\alpha \mu}^{\beta \nu}(v) =
  \begin{tikzpicture}[baseline=-.5ex,>=latex]
    \draw (0,-1) node[below]{$\alpha$} -- (0,1) node[above]{$\beta$};
    \draw (-1,0) node[anchor=mid east]{$\mu$} -- (1,0) node[right]{$\nu$};
    \draw[->] (0,0) -- (0,0.6);
    \draw[->] (0,0) -- (0.6,0);
    \node[above right=1pt] at (0,0) {$v$};
  \end{tikzpicture}$
  \caption{Graphical depiction of the $R$-matrix as defined in equation~\eqref{eq:rmatrix}.}
  \label{fig:vertex}
\end{figure}
The one-dimensional US model is known to be exactly solvable by Bethe
ansatz~(BA). The two-dimensional classical counterpart is given by the
Perk-Schultz~(PS) model~\cite{perk81}, which is defined on a square lattice
with $L \times N$ sites and periodic boundary conditions in both
directions. Variables taking on integer values from $1$ to $q$ are assigned to
each bond of the lattice, and a Boltzmann weight depending on a spectral
parameter $v$,
\begin{equation}\label{eq:rmatrix}
  R_{\alpha \mu}^{\beta \nu}(v) = \delta_{\alpha \nu} \delta_{\mu \beta} + v
  \cdot (-1)^{p(\alpha) p(\mu)} \cdot \delta_{\alpha \beta} \delta_{\mu \nu}\msp,
\end{equation}
is associated to every local vertex configuration $(\alpha, \beta, \mu, \nu)$,
see Figure~\vref{fig:vertex}. We define the row-to-row transfer matrix
\begin{equation}\label{eq:tmatrix}
  \opr T_\alpha^\beta(v) = \sum_{\{\nu\}} \prod_{j=1}^L R_{\alpha_j
  \nu_j}^{\beta_j \nu_{j+1}}(v)\msp.
\end{equation}
As the $R$-matrix~\eqref{eq:rmatrix} is a solution to the Yang-Baxter
equation, these transfer matrices form a commuting family $[\opr T(v), \opr
T(v')] = 0$ for all $v,v' \in \set{C}$. Making use of Baxter's
formula~\cite{baxter82}, one recovers the Hamiltonian of the US model from the
transfer matrix of the PS model at the shift point $v = 0$,
\begin{equation}\label{eq:threlation}
  \opr H_0 = \left. \frac{\dif}{\dif v} \ln\opr T(v) \right|_{v = 0} =
  \sum_{j=1}^L \pi_{j,j+1}\msp.
\end{equation}
Without breaking integrability, we may add external field terms,
\begin{equation}\label{eq:exthamilton}
  \opr H = \opr H_0 + \opr H_\text{ext} = \opr H_0 - \sum_{j=1}^L \sum_{\alpha
    = 1}^q \mu_\alpha n_{j,\alpha}\msp,
\end{equation}
where $\mu_\alpha$ is some general chemical potential associated with state
$\alpha$ and the operator $n_{j,\alpha}$ counts the number of particles of
type $\alpha$ sitting on site $j$.

We are interested in the thermodynamics of the US model. Hence, we want
to establish some connection between its partition function and the transfer
matrix of the PS model. We therefore consider a second $R$-matrix, namely
$\widebar R(v)$, obtained by rotating the graphical depiction of $R(v)$
clockwise by 90 degrees,
\begin{equation}
  \widebar R_{\alpha \mu}^{\beta \nu}(v) = R_{\nu \alpha}^{\mu \beta}(v)\msp.
\end{equation}
We define the transfer matrix $\widebar{\opr T}(v)$ as the product
of matrices $\widebar R(v)$ in analogy
to~\eqref{eq:tmatrix}. Equation~\eqref{eq:threlation} now also applies to
$\widebar{\opr T}(v)$, and as a consequence the relation
\begin{equation}
  \opr T(-\beta / N) \widebar{\opr T}(-\beta / N) = \ex{-2 (\beta / N)
  \opr H_0 + \mathcal O\left((\beta/N)^2\right)}
\end{equation}
is valid for arbitrary inverse temperature $\beta$ and a sufficiently
large even integer Trotter number $N$. The partition function of the
one-dimensional US model is then given by
\begin{equation}\label{eq:tpartition}
  Z = \Tr\ex{-\beta \opr H} = \lim_{N \to \infty} \Tr \left[\left(\opr T(u)
  \widebar{\opr T}(u)\right)^{N / 2} \ex{-\beta \opr H_\text{ext}}\right]\msp,
\end{equation}
where $u = -\beta / N$ and the traces are taken in the $q^L$-dimensional
space. Obviously, the partition function of the one-dimensional US model
is equal to the partition function of a staggered two-dimensional PS model,
where the external field can be incorporated by modifying the boundary
conditions in the Trotter direction. Equation~\eqref{eq:tpartition} is
still difficult to evaluate, e.g.\ all eigenstates have to be taken into
account. To avoid this problem, it is better to consider the column-to-column
transfer matrix of the staggered vertex model, which is called the quantum
transfer matrix~(QTM)~\cite{suzuki85,kluemper92b}. In order to write down the
QTM in a convenient way, we first define the matrix $\widetilde R(v)$, which
we get by rotating $R(v)$ counterclockwise by 90 degrees and changing the sign
of the spectral parameter,
\begin{equation}
  \widetilde R_{\alpha \mu}^{\beta \nu}(v) = R_{\mu \beta}^{\nu \alpha}(-v) \msp.
\end{equation}
Then the QTM takes the form
\begin{equation}
  \left(\QTM\right)_\alpha^\beta(v) = \sum_{\{\nu\}} \ex{\beta \mu_{\nu_1}} \prod_{j=1}^{N/2}
  R_{\alpha_{2j-1} \nu_{2j-1}}^{\beta_{2j-1} \nu_{2j}}(\im v + u) \widetilde
  R_{\alpha_{2j} \nu_{2j}}^{\beta_{2j} \nu_{2j+1}}(\im v - u)\msp,
\end{equation}
where we have introduced a new spectral parameter $v$, so that
the QTMs for all $v, v' \in \set{C}$ form a commuting family,
\begin{equation}
  \left[\QTM(v), \QTM(v')\right] = 0\msp.
\end{equation}
This allows for the diagonalisation by use of the BA. In the end, we are only
interested in the case $v = 0$ as the partition function
of the one-dimensional US model in terms of the QTM is
\begin{equation}\label{eq:partqtm}
  Z = \lim_{N \to\infty} \Tr \left(\QTM(0)\right)^L\msp.
\end{equation}
In the thermodynamic limit ($L \to \infty$) one finds that the thermodynamics
of the US model solely depends on the unique largest eigenvalue of the
QTM~\cite{suzuki87,suzuki90}. For the free energy per unit length, we finally
get
\begin{equation}\label{eq:energy}
  f = -\lim_{L \to \infty} \frac{1}{L \beta} \ln Z = -\frac{1}{\beta} \ln \Lambda_\text{max}(0)\msp,
\end{equation}
where $\Lambda_\text{max}(v)$ is the largest eigenvalue of the QTM.

As noted before, the QTM can be diagonalised via BA. The result for the
eigenvalue is~\cite{kluemper97}
\begin{equation}\label{eq:eigenvalue}
  \Lambda(v) = \sum_{j=1}^q \lambda_j(v)\msp,
\end{equation}
where
\begin{equation}
  \lambda_j(v) = \phi_-(v) \phi_+(v) \frac{q_{j-1}(v -
  \im\epsilon_j)}{q_{j-1}(v)} \frac{q_j(v + \im\epsilon_j)}{q_j(v)} \ex{\beta\mu_j}\msp.
\end{equation}
For convenience, we have defined the functions $\phi_\pm(v) = (v \pm \im u)^{N/2}$ and
\begin{equation}\label{eq:qfunct}
  q_j(v) = \begin{cases}
  \phi_-(v) &\text{for $j = 0$}\\
  \prod_{k_j=1}^{M_j} (v - \vbethe{j}) & \text{for $j = 1,\ldots,q-1$}\\
  \phi_+(v) &\text{for $j = q$}\end{cases}\msp,
\end{equation}
where the complex parameters $\vbethe{j}$ are the so-called BA roots and $M_j$
is the total number of BA roots
in set $j$. The BA roots have to fulfil the BA equations
\begin{equation}\label{eq:baeq}
  \frac{\lambda_j(\vbethe{j})}{\lambda_{j+1}(\vbethe{j})} = -1\msp,
\end{equation}
to ensure that all potential poles in the expression~\eqref{eq:eigenvalue} for
$\Lambda(v)$, which has to be a polynomial of degree $N$, cancel. The BA
equations form a system of coupled nonlinear equations for the unknown BA
roots.

\section{Nonlinear integral equations for the \texorpdfstring{$\bm{\alg{sl}(4)}$}{sl(4)} case}\label{sec:nliesl4}

Let us now turn to the special case of the $\alg{sl}(4)$-symmetric
Uimin-Sutherland model ($q=4$, $\epsilon_j = +1$ for all $j$). The largest
eigenvalue of the QTM lies in the sector $M_j = N / 2$ for all $j$. Of course,
it is in principle possible to solve the BA equations~\eqref{eq:baeq} for some
fixed Trotter number $N$. But this approach is possible only for finite $N$
and is also quite cumbersome to do numerically. As we are interested in the
limit $N \to \infty$ for deriving the free energy of the model, we have to
encode the BA equations into a form for which this limit can be taken
analytically.

We start by defining some suitable auxiliary functions, which will in the end
turn out to fulfil certain nonlinear integral equations. For convenience, we
will use an abbreviated notation utilising the Yangian analogue of
Young tableaux~\cite{bazhanov90,suzuki94,kuniba95a,kuniba95b}. Instead of the
function $\lambda_j(v)$, we will write a box filled with the letter $j$,
\begin{equation}
  \young(j) = \lambda_j(v)\msp.
\end{equation}
This corresponds to a Young tableau belonging to a vector of the
first, four-dimensional fundamental representation of $\alg{sl}(4)$. We also
define a Young tableau belonging to the second, six-dimensional
representation,
\begin{align}
  \young(j,k) = \lambda_j(v - \im/2) \lambda_k(v + \im/2)\msp,
\end{align}
and for the conjugate four-dimensional representation we have
\begin{equation}
  \young(j,k,l) = \lambda_j(v - \im) \lambda_k(v) \lambda_l(v + \im)\msp.
\end{equation}
From fusion hierarchy~\cite{kuniba94,juettner98} one knows that the
eigenvalues of the QTMs belonging to the three fundamental representations can
be written as
\begin{subequations}\label{eq:fundeigenv}
  \begin{align}
    \rep{\Lambda}{1}(x) &= \left.\young(1) + \young(2) + \young(3) + \young(4)\right|_{v=x}\msp,\label{eq:eigenvrep1}\\
    \rep{\Lambda}{2}(x) &= \left.\young(1,2) + \young(1,3) + \young(1,4)
      + \young(2,3) + \young(2,4) + \young(3,4)\right|_{v=x}\msp,\\
    \rep{\Lambda}{3}(x) &= \left.\young(1,2,3) + \young(1,2,4)
      + \young(1,3,4) + \young(2,3,4)\right|_{v=x}\msp,
  \end{align}
\end{subequations}
where the superscripts denote the representations. In all three cases
the BA equations~\eqref{eq:baeq} ensure that the eigenvalues are free of
poles. We note that~\eqref{eq:eigenvrep1}
is equivalent to~\eqref{eq:eigenvalue}. We define the following four auxiliary
functions for the first fundamental representation:
\begin{subequations}\label{eq:aux1}
  \begin{align}
    \rep{b}{1}_1(x) &= \left.\frac{\young(1)}{\young(2) + \young(3) +
        \young(4)}\right|_{v = x + \im/2}\msp,\\
    \rep{b}{1}_2(x) &= \left.\frac{\young(1,2) \cdot \left(\young(2,3) + \young(2,4) +
          \young(3,4)\right)}{\left(\young(1,3) + \young(1,4)\right) \cdot \left(\young(1,2)
          + \young(1,3) + \young(1,4) + \young(2,3) + \young(2,4) +
          \young(3,4)\right)}\right|_{v = x}\msp,\\
    \rep{b}{1}_3(x) &= \left.\frac{\young(1,3) \cdot \young(3,4)}{\young(1,4) \cdot
        \left(\young(1,3) + \young(1,4) + \young(2,3) + \young(2,4) +
          \young(3,4)\right)}\right|_{v = x}\msp,\\
    \rep{b}{1}_4(x) &= \left.\frac{\young(4)}{\young(1) + \young(2) +
        \young(3)}\right|_{v = x - \im/2}\msp.
  \end{align}
\end{subequations}
We have six auxiliary functions for the second fundamental representation:
\begin{subequations}\label{eq:aux2}
  \begin{align}
    \rep{b}{2}_1(x) &= \left.\frac{\young(1,2)}{\young(1,3) + \young(1,4) +
        \young(2,3) + \young(2,4) + \young(3,4)}\right|_{v = x + \im/2}\msp,\\
    \rep{b}{2}_2(x) &= \left.\frac{\young(1,3) \cdot \young(3,4)}{\left(\young(1,4)
          + \young(2,4) + \young(3,4)\right) \cdot \left(\young(2,3) + \young(2,4)
          + \young(3,4)\right)}\right|_{v = x + \im/2}\msp,\\
    \rep{b}{2}_3(x) &= \left.\frac{\young(1) \cdot \young(4)}{\left(\young(2) +
          \young(3)\right) \cdot \left(\young(1) + \young(2) + \young(3) +
          \young(4)\right)}\right|_{v = x}\msp,\\
    \rep{b}{2}_4(x) &= \left.\frac{\young(1,2,3) \cdot
        \young(2,3,4)}{\left(\young(1,2,4) + \young(1,3,4)\right) \cdot
        \left(\young(1,2,3) + \young(1,2,4) +
          \young(1,3,4) + \young(2,3,4)\right)}\right|_{v = x}\msp,\\
    \rep{b}{2}_5(x) &= \left.\frac{\young(1,2) \cdot \young(2,4)}{\left(\young(1,2)
          + \young(1,3) + \young(1,4)\right) \cdot \left(\young(1,2) + \young(1,3)
          + \young(2,3)\right)}\right|_{v = x - \im/2}\msp,\\
    \rep{b}{2}_6(x) &= \left.\frac{\young(3,4)}{\young(1,2) + \young(1,3) +
        \young(1,4) + \young(2,3) + \young(2,4)}\right|_{v = x - \im/2}\msp.
  \end{align}
\end{subequations}
And finally, the four auxiliary functions for the third fundamental
representation are:
\begin{subequations}\label{eq:aux3}
  \begin{align}
    \rep{b}{3}_1(x) &= \left.\frac{\young(1,2,3)}{\young(1,2,4) + \young(1,3,4) +
        \young(2,3,4)}\right|_{v = x + \im/2}\msp,\\
    \rep{b}{3}_2(x) &= \left.\frac{\young(1,2) \cdot \young(2,4)}{\young(2,3) \cdot
        \left(\young(1,2) + \young(1,3) + \young(1,4) + \young(2,3) +
          \young(2,4)\right)}\right|_{v = x}\msp,\\
    \rep{b}{3}_3(x) &= \left.\frac{\young(3,4) \cdot \left(\young(1,2) + \young(1,3) +
          \young(1,4)\right)}{\left(\young(2,3) + \young(2,4)\right) \cdot
        \left(\young(1,2) + \young(1,3) + \young(1,4) + \young(2,3) +
          \young(2,4) + \young(3,4)\right)}\right|_{v = x}\msp,\\
    \rep{b}{3}_4(x) &= \left.\frac{\young(2,3,4)}{\young(1,2,3) + \young(1,2,4) +
        \young(1,3,4)}\right|_{v = x - \im/2}\msp.
  \end{align}
\end{subequations}
In addition to the auxiliary functions above, we define a second set of
functions, namely $\rep{B}{n}_j(x) = \rep{b}{n}_j(x) + 1$. These can also be
written in a form, where only simple sums of Young tableaux appear as factors
in the numerators and denominators, and all factors appearing
in the set of functions $\rep{b}{n}_j(x)$ also appear in the set of functions
$\rep{B}{n}_j(x)$. We note that all these factors are
partial sums of the Young tableaux appearing in the
eigenvalues~\eqref{eq:fundeigenv}. The set of functions~\eqref{eq:aux1} is
related to~\eqref{eq:aux3} by a conjugation transformation. The
set~\eqref{eq:aux2} is self-conjugate in this sense. Furthermore, all
auxiliary functions are rational functions in terms of the spectral parameter
$x$ and are analytic, non-zero and have constant asymptotics~(ANZC) in a strip
$-1/2 \lesssim \Im(x) \lesssim 1/2$.

Now, the actual calculation is rather straightforward but lengthy. Therefore,
details are deferred to Appendix~\ref{app:nliederiv}. The key idea is to
apply a Fourier transform to the logarithmic derivative of all auxiliary
functions and to exploit their analyticity properties in Fourier space. We
like to stress that, although we are working with arbitrary finite $N$
throughout the derivation, in the end the limit $N \to \infty$ can be taken
analytically.

We finally arrive at a system of coupled nonlinear integral equations~(NLIE) of
the form
\begin{equation}\label{eq:nlie}
  \vecl b(x) = -\beta \vect\epsilon(x) - \bigl[\matr K \ast \vecl B\bigr](x)\msp,
\end{equation}
where we have defined
\begin{align}
  \vecl b &= \left(\ln\rep{b}{1}_1, \ldots, \ln\rep{b}{1}_4,
    \ln\rep{b}{2}_1, \ldots, \ln\rep{b}{2}_6, \ln\rep{b}{3}_1, \ldots,
    \ln\rep{b}{3}_4\right)^\mathrm T\msp,\\
  \vecl B &= \left(\ln\rep{B}{1}_1, \ldots, \ln\rep{B}{1}_4, \ln\rep{B}{2}_1,
    \ldots, \ln\rep{B}{2}_6, \ln\rep{B}{3}_1, \ldots,
    \ln\rep{B}{3}_4\right)^\mathrm T\msp,\\
  \vect\epsilon &= \left(\rep{\epsilon}{1}_1, \ldots, \rep{\epsilon}{1}_4,
    \rep{\epsilon}{2}_1, \ldots, \rep{\epsilon}{2}_6, \rep{\epsilon}{3}_1,
    \ldots, \rep{\epsilon}{3}_4\right)^\mathrm T\msp.\label{eq:bareenergy}
\end{align}
Convolutions are denoted by
\begin{equation}
  \bigl[f \ast g\bigr](x) = \int_{-\infty}^\infty f(x-y) g(y)\,\frac{\dif y}{2 \pi}\msp.
\end{equation}
The kernel matrix $\matr K(x)$ is a 14 by 14 matrix. As this is too
large to be displayed as a whole, we divide the matrix into blocks
connecting the auxiliary functions from different representations,
\begin{equation}\label{eq:nliekernel}
  \matr K(x) =
  \begin{pmatrix}
    \rep{\matr K}{1,1}(x) & \rep{\matr K}{1,2}(x) & \rep{\matr K}{1,3}(x)\\
    \rep{\matr K}{2,1}(x) & \rep{\matr K}{2,2}(x) & \rep{\matr K}{2,3}(x)\\
    \rep{\matr K}{3,1}(x) & \rep{\matr K}{3,2}(x) & \rep{\matr K}{3,3}(x)
  \end{pmatrix}\msp.
\end{equation}
Obviously, $\rep{\matr K}{1,1}(x)$ is a 4 by 4 matrix, while $\rep{\matr K}{1,2}(x)$ is a
4 by 6 matrix, etc. $\matr K(x)$ is Hermitian and invariant under
reflection along the anti-diagonal,
\begin{align}
  \matr K(x) &= \bigl[\matr K(x)\bigr]^\dagger\msp,
  & \bigl[\matr K(x)\bigr]_{j,k} &= \bigl[\matr K(x)\bigr]_{15-k,15-j}\msp.
\end{align}
Therefore we only need to consider
\begin{subequations}
  \begin{align}
    \rep{\matr K}{1,1}(x) &=
    \begin{pmatrix}
      K_0(x) & K_1(x) & K_1(x) & K_1(x)\\
      K_2(x) & K_0(x) & K_1(x) & K_1(x)\\
      K_2(x) & K_2(x) & K_0(x) & K_1(x)\\
      K_2(x) & K_2(x) & K_2(x) & K_0(x)
    \end{pmatrix}\msp,\\
    \rep{\matr K}{1,2}(x) &=
    \begin{pmatrix}
      K_{11}(x) & K_{11}(x) & K_{11}(x) & K_{12}(x) & K_{12}(x) & K_{12}(x)\\
      K_{11}(x) & K_{14}(x) & K_{14}(x) & K_{11}(x) & K_{11}(x) & K_{12}(x)\\
      K_{13}(x) & K_{11}(x) & K_{14}(x) & K_{11}(x) & K_{14}(x) & K_{11}(x)\\
      K_{13}(x) & K_{13}(x) & K_{11}(x) & K_{13}(x) & K_{11}(x) & K_{11}(x)
    \end{pmatrix}\msp,\\
    \rep{\matr K}{1,3}(x) &=
    \begin{pmatrix}
      K_{15}(x) & K_{15}(x) & K_{15}(x) & K_{16}(x)\\
      K_{15}(x) & K_{15}(x) & K_{18}(x) & K_{15}(x)\\
      K_{15}(x) & K_{19}(x) & K_{15}(x) & K_{15}(x)\\
      K_{17}(x) & K_{15}(x) & K_{15}(x) & K_{15}(x)
    \end{pmatrix}\msp,\\
    \rep{\matr K}{2,2}(x) &=
    \begin{pmatrix}
      K_3(x) & K_4(x) & K_4(x) & K_4(x) & K_4(x) & K_6(x)\\
      K_5(x) & K_3(x) & K_4(x) & K_4(x) & K_8(x) & K_4(x)\\
      K_5(x) & K_5(x) & K_3(x) & K_{10}(x) & K_4(x) & K_4(x)\\
      K_5(x) & K_5(x) & K_{10}(x) & K_3(x) & K_4(x) & K_4(x)\\
      K_5(x) & K_9(x) & K_5(x) & K_5(x) & K_3(x) & K_4(x)\\
      K_7(x) & K_5(x) & K_5(x) & K_5(x) & K_5(x) & K_3(x)
    \end{pmatrix}\msp.
  \end{align}
\end{subequations}
The remaining matrices easily follow from the relations
\begin{subequations}
  \begin{align}
    \rep{\matr K}{3,3}(x) &= \rep{\matr K}{1,1}(x)\msp,
    & \rep{\matr K}{2,1}(x) &= \bigl[\rep{\matr K}{1,2}(x)\bigr]^\dagger\msp,\\
    \rep{\matr K}{3,1}(x) &= \bigl[\rep{\matr K}{1,3}(x)\bigr]^\dagger\msp,
    & \rep{\matr K}{3,2}(x) &= \bigl[\rep{\matr K}{2,3}(x)\bigr]^\dagger\msp,\\
    \bigl[\rep{\matr K}{2,3}(x)\bigr]_{j,k} &= \bigl[\rep{\matr
      K}{1,2}(x)\bigr]_{7-k,5-j}\msp.
  \end{align}
\end{subequations}
The kernels are defined as $K_j(x) = \int_{-\infty}^{\infty} \widehat
K_j(k) \ex{\im k x}\,\dif k$, where
\begin{subequations}
  \begin{align}
    \widehat K_0(k) &= \rep{\widehat K}{1,1}_{[4]}(k)\msp, &
    \widehat K_1(k) &= \rep{\widehat K}{1,1}_{[4]}(k) + \ex{-k/2 - |k|/2}\msp,\\
    \widehat K_2(k) &= \rep{\widehat K}{1,1}_{[4]}(k) + \ex{k/2 - |k|/2}\msp, &
    \widehat K_3(k) &= \rep{\widehat K}{2,2}_{[4]}(k)\msp,\\
    \widehat K_4(k) &= \rep{\widehat K}{2,2}_{[4]}(k) + \ex{-k/2 - |k|/2}\msp, &
    \widehat K_5(k) &= \rep{\widehat K}{2,2}_{[4]}(k) + \ex{k/2 - |k|/2}\msp,\\
    \widehat K_6(k) &= \rep{\widehat K}{2,2}_{[4]}(k) + \ex{-k - |k|}\msp, &
    \widehat K_7(k) &= \rep{\widehat K}{2,2}_{[4]}(k) + \ex{k - |k|}\msp,\\
    \widehat K_8(k) &= \rep{\widehat K}{2,2}_{[4]}(k) + 2 \ex{-k/2 - |k|/2}\msp, &
    \widehat K_9(k) &= \rep{\widehat K}{2,2}_{[4]}(k) + 2 \ex{k/2 - |k|/2}\msp,\\
    \widehat K_{10}(k) &= \rep{\widehat K}{2,2}_{[4]}(k) + \ex{-|k|}\msp, &
    \widehat K_{11}(k) &= \rep{\widehat K}{1,2}_{[4]}(k)\msp,\\
    \widehat K_{12}(k) &= \rep{\widehat K}{1,2}_{[4]}(k) + \ex{-k - |k|/2} - \ex{-k/2}\msp, &
    \widehat K_{13}(k) &= \rep{\widehat K}{1,2}_{[4]}(k) + \ex{k - |k|/2} - \ex{k/2}\msp,\\
    \widehat K_{14}(k) &= \rep{\widehat K}{1,2}_{[4]}(k) + \ex{-|k|/2}\msp, &
    \widehat K_{15}(k) &= \rep{\widehat K}{1,3}_{[4]}(k)\msp,\\
    \widehat K_{16}(k) &= \rep{\widehat K}{1,3}_{[4]}(k) + \ex{-3 k/2 - |k|/2} - \ex{-k}\msp, &
    \widehat K_{17}(k) &= \rep{\widehat K}{1,3}_{[4]}(k) + \ex{3 k/2 -|k|/2} - \ex{k}\msp,\\
    \widehat K_{18}(k) &= \rep{\widehat K}{1,3}_{[4]}(k) + \ex{-k/2 - |k|/2} - 1\msp, &
    \widehat K_{19}(k) &= \rep{\widehat K}{1,3}_{[4]}(k) + \ex{k/2 -|k|/2}\msp,
  \end{align}
\end{subequations}
with the function
\begin{equation}
  \rep{\widehat K}{n,m}_{[q]}(k) = \ex{|k|/2} \frac{\sinh(\min(n,m) k/2) \sinh([q-\max(n,m)]
    k/2)}{\sinh(k/2) \sinh(q k/2)} - \delta_{nm}\msp.
\end{equation}
We note that in spectral parameter space all kernels can be written in terms of digamma
and simple rational functions. Nevertheless our notation is more useful here as the
numerical treatment of the NLIE can be conveniently done in Fourier
space. The functions $\rep{K}{n,m}_{[q]}(x) =
\int_{-\infty}^\infty \rep{\widehat K}{n,m}_{[q]}(k) \ex{\im k x}\,\dif k$ are
related to the $S$-matrix of elementary excitations~\cite{kulish81} via
\begin{equation}
  \rep{K}{n,m}_{[q]}(x) = \frac{\dif}{\dif x}[\im \ln\rep{S}{n,m}_{[q]}(x)]\msp.
\end{equation}
The bare energies in~\eqref{eq:bareenergy} are $\rep{\epsilon}{n}_j(x) = \rep{V}{n}_{[4]}(x) +
\rep{c}{n}_j$, where
\begin{equation}\label{eq:nlievdriv}
  \rep{V}{n}_{[q]}(x) = \frac{2 \pi}{q} \frac{\sin(\pi n / q)}{\cosh(2 \pi x / q) -
  \cos(\pi n / q)}\msp,
\end{equation}
and the constants are given by
\begin{subequations}\label{eq:nlieconstants}
  \begin{align}
    \rep{c}{1}_1 &= (-3 \mu_1 + \mu_2 + \mu_3 + \mu_4)/4\msp, &
    \rep{c}{1}_2 &= ( \mu_1 - 3 \mu_2 + \mu_3 + \mu_4)/4\msp,\\
    \rep{c}{1}_3 &= ( \mu_1 + \mu_2 - 3 \mu_3 + \mu_4)/4\msp, &
    \rep{c}{1}_4 &= ( \mu_1 + \mu_2 + \mu_3 - 3 \mu_4)/4\msp,\\
    \rep{c}{2}_1 &= (-\mu_1 - \mu_2 + \mu_3 + \mu_4)/2\msp, &
    \rep{c}{2}_2 &= (-\mu_1 + \mu_2 - \mu_3 + \mu_4)/2\msp,\\
    \rep{c}{2}_3 &= (-\mu_1 + \mu_2 + \mu_3 - \mu_4)/2\msp, &
    \rep{c}{2}_4 &= ( \mu_1 - \mu_2 - \mu_3 + \mu_4)/2\msp,\\
    \rep{c}{2}_5 &= ( \mu_1 - \mu_2 + \mu_3 - \mu_4)/2\msp, &
    \rep{c}{2}_6 &= ( \mu_1 + \mu_2 - \mu_3 - \mu_4)/2\msp,\\
    \rep{c}{3}_1 &= (-\mu_1 - \mu_2 - \mu_3 + 3 \mu_4)/4\msp, &
    \rep{c}{3}_2 &= (-\mu_1 - \mu_2 + 3 \mu_3 - \mu_4)/4\msp,\\
    \rep{c}{3}_3 &= (-\mu_1 + 3 \mu_2 - \mu_3 - \mu_4)/4\msp, &
    \rep{c}{3}_4 &= ( 3 \mu_1 - \mu_2 - \mu_3 - \mu_4)/4\msp.
  \end{align}
\end{subequations}

Finally, the largest eigenvalue of the QTM can be written in terms of the auxiliary
functions,
\begin{equation}\label{eq:nlielambda}
  \ln\Lambda_\text{max}(0) = -\beta \biggl(1 - \frac{\pi}{4} - \frac{3}{2} \ln 2 - \frac{1}{4} \sum_{j=1}^4 \mu_j\biggr) +
  \sum_{n=1}^3 \sum_{j=1}^{d_n} \left[\rep{V}{n}_{[4]} \ast \ln\rep{B}{n}_j\right](0)\msp,
\end{equation}
where $d_n = \binom{4}{n}$ is the dimension of the $n$th fundamental representation.
Therefore, the problem of solving the infinitely many BA equations~\eqref{eq:baeq} in the limit $N \to \infty$
has been reduced to finding a finite set of functions satisfying the
NLIE~\eqref{eq:nlie}--\eqref{eq:nlieconstants}. The NLIE is valid for
arbitrary finite temperature and chemical potentials.

\section{Analytical investigation}\label{sec:analytical}

\subsection{Investigation of the \texorpdfstring{$\bm{\alg{sl}(3)}$}{sl(3)} limit}

First we want to show how our formulation~\eqref{eq:nlie}--\eqref{eq:nlielambda} reduces to the known NLIE
for the $\alg{sl}(3)$-symmetric case~\cite{fujii99} by freezing out one of the
states. We choose the state $\alpha = 4$ and accordingly treat the limit
$\mu_4 \to -\infty$. We observe that
\begin{subequations}
  \begin{align}
    \rep{b}{1}_1(x) &= \ord(1)\msp, &
    \rep{b}{1}_2(x) &= \ord(1)\msp, &
    \rep{b}{1}_3(x) &= \ord(1)\msp,\\
    \rep{b}{1}_4(x) &= \ord(\ex{\beta\mu_4})\msp, &
    \rep{b}{2}_1(x) &= \ord(1)\msp, &
    \rep{b}{2}_2(x) &= \ord(1)\msp,\\
    \rep{b}{2}_3(x) &= \ord(\ex{\beta\mu_4})\msp, &
    \rep{b}{2}_4(x) &= \ord(1)\msp, &
    \rep{b}{2}_5(x) &= \ord(\ex{\beta\mu_4})\msp,\\
    \rep{b}{2}_6(x) &= \ord(\ex{\beta\mu_4})\msp, &
    \rep{b}{3}_1(x) &= \ord(\ex{-\beta\mu_4})\msp, &
    \rep{b}{3}_2(x) &= \ord(\ex{\beta\mu_4})\msp,\\
    \rep{b}{3}_3(x) &= \ord(\ex{\beta\mu_4})\msp, &
    \rep{b}{3}_4(x) &= \ord(\ex{\beta\mu_4})\msp.
  \end{align}
\end{subequations}
Therefore only seven of the auxiliary functions survive. We can regard
\begin{equation}
  \rep{b}{1}_4(x) \equiv \rep{b}{2}_3(x) \equiv \rep{b}{2}_5(x) \equiv \rep{b}{2}_6(x) \equiv
  \rep{b}{3}_2(x) \equiv \rep{b}{3}_3(x) \equiv \rep{b}{3}_4(x) \equiv 0\msp.
\end{equation}
We also conclude that $\rep{b}{3}_1(x) / \rep{B}{3}_1(x) \to 1$. Using this information, the equation for
$\ln\rep{b}{3}_1(x)$ linearises and can be solved analytically. We get
\begin{multline}
  \ln\rep{B}{3}_1(x) = -\beta \left(W(x) -
    \frac{\mu_1+\mu_2+\mu_3}{3} + \mu_4\right)\\
  - \left[\rep{V}{2}_{[3]} \ast \ln(\rep{B}{1}_1 \rep{B}{1}_2
  \rep{B}{1}_3)\right](x) -\left[\rep{V}{1}_{[3]} \ast \ln(\rep{B}{2}_1 \rep{B}{2}_2
  \rep{B}{2}_4)\right](x)\msp,
\end{multline}
where
\begin{equation}
W(x) = \int_{-\infty}^\infty \frac{\ex{-|k|/2}}{\ex{-k}+1+\ex{k}}
  \ex{\im k x}\,\dif k\msp.
\end{equation}
Substituting this into our NLIE and relabelling $\rep{b}{2}_4(x)$ to $\rep{b}{2}_3(x)$, we are again left with a
NLIE of type~\eqref{eq:nlie}, but with only six remaining auxiliary functions
belonging to the two fundamental representations of $\alg{sl}(3)$. Here we get
the kernel matrix
\begin{equation}
  \matr K(x) =
  \begin{pmatrix}
    \rep{\matr K}{1,1}(x) & \rep{\matr K}{1,2}(x)\\
    \rep{\matr K}{1,2}(x) & \rep{\matr K}{1,1}(x)
  \end{pmatrix}\msp,
\end{equation}
where the submatrices are given by
\begin{subequations}
  \begin{align}
    \rep{\matr K}{1,1}(x) &=
    \begin{pmatrix}
      K_0(x) & K_1(x) & K_1(x)\\
      K_2(x) & K_0(x) & K_1(x)\\
      K_2(x) & K_2(x) & K_0(x)
    \end{pmatrix}\msp,\\
    \rep{\matr K}{1,2}(x) &=
    \begin{pmatrix}
      K_3(x) & K_3(x) & K_4(x)\\
      K_3(x) & K_6(x) & K_3(x)\\
      K_5(x) & K_3(x) & K_3(x)
    \end{pmatrix}\msp.
  \end{align}
\end{subequations}
The Fourier transformed kernels are found to be
\begin{subequations}
  \begin{align}
    \widehat K_0(k) &= \rep{\widehat K}{1,1}_{[3]}(k)\msp, &
    \widehat K_1(k) &= \rep{\widehat K}{1,1}_{[3]}(k) + \ex{-k/2 - |k|/2}\msp,\\
    \widehat K_2(k) &= \rep{\widehat K}{1,1}_{[3]}(k) + \ex{k/2 - |k|/2}\msp, &
    \widehat K_3(k) &= \rep{\widehat K}{1,2}_{[3]}(k)\msp,\\
    \widehat K_4(k) &= \rep{\widehat K}{1,2}_{[3]}(k) + \ex{-k - |k|/2} - \ex{-k/2}\msp, &
    \widehat K_5(k) &= \rep{\widehat K}{1,2}_{[3]}(k) + \ex{k - |k|/2} - \ex{k/2}\msp,\\
    \widehat K_6(k) &= \rep{\widehat K}{1,2}_{[3]}(k) + \ex{-|k|/2}\msp.
  \end{align}
\end{subequations}
For the bare energies, we get
$\rep{\epsilon}{n}_j(x) = \rep{V}{n}_{[3]}(x) + \rep{c}{n}_j$ with the constants
\begin{subequations}\label{eq:nliesl3const}
  \begin{align}
    \rep{c}{1}_1 &= (-2 \mu_1 + \mu_2 + \mu_3) / 3\msp, &
    \rep{c}{1}_2 &= ( \mu_1 - 2 \mu_2 + \mu_3) / 3\msp,\\
    \rep{c}{1}_3 &= ( \mu_1 + \mu_2 - 2 \mu_3) / 3\msp, &
    \rep{c}{2}_1 &= (-\mu_1 - \mu_2 + 2 \mu_3) / 3\msp,\\
    \rep{c}{2}_2 &= (-\mu_1 + 2 \mu_2 - \mu_3) / 3\msp, &
    \rep{c}{2}_3 &= ( 2 \mu_1 - \mu_2 - \mu_3) / 3\msp.
  \end{align}
\end{subequations}
The largest eigenvalue is finally given by
\begin{equation}
  \ln\Lambda_\text{max}(0) = -\beta \biggl(1 - \frac{\pi}{3 \sqrt{3}} - \ln 3 - \frac{1}{3} \sum_{j=1}^3 \mu_j\biggr) +
  \sum_{n=1}^2 \sum_{j=1}^{3} \left[\rep{V}{n}_{[3]} \ast
  \ln\rep{B}{n}_j\right](0)\msp.
\end{equation}
As expected, this is exactly the known NLIE for the $\alg{sl}(3)$-symmetric
case~\cite{fujii99}.

We can also recover the explicit form of all auxiliary functions for the
$\alg{sl}(3)$ case. We drop all Young tableaux that contain
$\young(4)$ in the auxiliary functions~\eqref{eq:aux1}--\eqref{eq:aux3} as $\lambda_4(v)
\to 0$ in the limit $\mu_4 \to -\infty$. We have already seen that seven
of the functions become zero, while one function diverges. After relabelling
$\rep{b}{2}_4(x)$ to $\rep{b}{2}_3(x)$ again, the remaining six
functions take the form
\begin{subequations}\label{eq:sl3funct}
  \begin{align}
    \rep{b}{1}_1(x) &= \left.\frac{\young(1)}{\young(2) +
        \young(3)}\right|_{v = x + \im/2}\msp, &
    \rep{b}{1}_2(x) &= \left.\frac{\young(1,2) \cdot
        \young(2,3)}{\young(1,3) \cdot \left(\young(1,2) + \young(1,3) +
          \young(2,3)\right)}\right|_{v = x}\msp,\\
    \rep{b}{1}_3(x) &= \left.\frac{\young(3)}{\young(1) +
        \young(2)}\right|_{v = x - \im/2}\msp, &
    \rep{b}{2}_1(x) &= \left.\frac{\young(1,2)}{\young(1,3) +
        \young(2,3)}\right|_{v = x + \im/2}\msp,\\
    \rep{b}{2}_2(x) &= \left.\frac{\young(1) \cdot
        \young(3)}{\young(2) \cdot \left(\young(1) + \young(2) +
          \young(3)\right)}\right|_{v = x}\msp, &
    \rep{b}{2}_3(x) &= \left.\frac{\young(2,3)}{\young(1,2) +
        \young(1,3)}\right|_{v = x - \im/2}\msp.
    \end{align}
\end{subequations}
Although only two sets of BA roots are present in the $\alg{sl}(3)$ case,
these auxiliary functions still contain the function $q_3(v)$, which enters
through $\lambda_3(v)$. But, as already indicated by the definition of
$q_3(v)$ for the case $q = 3$ in~\eqref{eq:qfunct}, we have to demand
$q_3(v) = \phi_+(v)$ in the limit $\mu_4 \to -\infty$ in order that the
previously derived NLIE can also be derived directly from the auxiliary
functions~\eqref{eq:sl3funct}. The auxiliary functions are equal to those
presented in~\cite{fujii99}.

We finally note that our choice of freezing out the state $\alpha = 4$ is
completely arbitrary. Choosing one of the other states yields, after
relabelling some of the indices, the same NLIE and auxiliary functions. For
$\mu_1 \to -\infty$, we find that $q_1(v) = \phi_-(v)$, while $q_2(v)$ and
$q_3(v)$ contain the remaining BA roots. For $\mu_\alpha \to -\infty$ with
$\alpha = 2$ or $3$, we find that $q_{\alpha - 1}(v) = q_\alpha(v)$.

\subsection{Limit \texorpdfstring{$\bm{T \to 0}$}{T to zero} and critical fields}

We divide the NLIE~\eqref{eq:nlie} by $\beta$ and define rescaled
auxiliary functions by
\begin{align}
  \rep{e}{n}_j(x) &= \frac{1}{\beta} \ln\rep{b}{n}_j(x)\msp, & \rep{E}{n}_j(x) &= \frac{1}{\beta} \ln\rep{B}{n}_j(x)\msp.
\end{align}
In the limit $T \to 0$ ($\beta \to \infty$) we get
\begin{equation}
  \rep{E}{n}_j(x) \to \rep{e^+}{n}_j(x) =
  \begin{cases}
    \rep{e}{n}_j(x) & \text{if $\Re(\rep{e}{n}_j(x)) > 0$}\\
    0 & \text{if $\Re(\rep{e}{n}_j(x)) \leq 0$}
  \end{cases}\msp.
\end{equation}
Obviously, auxiliary functions with negative real parts for all $x \in
\set{R}$ do no longer contribute to the ground-state energy, because
$\rep{e^+}{n}_j(x) \equiv 0$ for these functions.

Without loss of generality, we choose the chemical potentials to be ordered,
$\mu_1 \geq \mu_2 \geq \mu_3 \geq \mu_4$. Changing the order just amounts to
some permutation of indices in the following calculations. In our case, we observe that
the only remaining auxiliary functions are $\rep{e}{n}_1(x)$ for $n = 1,2,3$,
i.e.\ one from every representation. The NLIE linearise and take the form
\begin{equation}\label{eq:nlielin}
  \rep{e}{n}_1(x) = -\rep{V}{n}_{[4]}(x) - \rep{c}{n}_1 - \sum_{m=1}^{3}
  \bigl[\rep{K}{n,m}_{[4]} \ast \rep{e^+}{m}_1\bigr](x)\qquad(n = 1,2,3)\msp.
\end{equation}
It follows that the remaining auxiliary functions are real and symmetric with
respect to the spectral parameter. The ground-state energy is given by
\begin{equation}\label{eq:nlielineigenv}
  f_0 = 1 - \frac{\pi}{4} - \frac{3}{2} \ln 2 - \frac{1}{4} \sum_{j=1}^4 \mu_j -
  \sum_{n=1}^3 \bigl[\rep{V}{n}_{[4]} \ast \rep{e^+}{n}_1\bigr](0)\msp.
\end{equation}
These equations have a particularly simple solution if all chemical
potentials are equal, $\mu_1 = \mu_2 = \mu_3 = \mu_4$. In this case, we get
$\rep{e}{n}_1(x) = -\rep{V}{n}_{[4]}(x)$ and $\rep{e^+}{n}_1(x) \equiv 0$ for
all $n$. Therefore, the ground-state energy is just $f_0 = 1 - \pi / 4 - 3 \ln
(2) / 2$.

In general, depending on certain differences of the chemical potentials, the
ground state can be in one of four possible phases. We start with the phase,
where all degrees of freedom are frozen out, i.e.\ only the state $\alpha = 1$
survives. In this case, we have $\rep{e}{n}_1(x) = \rep{e^+}{n}_1(x)$ for all
$n$. As a consequence, \eqref{eq:nlielin} can be solved
analytically. We find the restriction $\mu_1 - \mu_2 \geq 4$ and obtain
\begin{align}
  \rep{e}{1}_1(x) &= \mu_1 - \mu_2 - \frac{4}{4 x^2 + 1}\msp, &
  \rep{e}{2}_1(x) &= \mu_2 - \mu_3\msp, & \rep{e}{3}_1(x) &= \mu_3 - \mu_4\msp,
\end{align}
while the ground-state energy turns out to be $f_0 = 1 - \mu_1$. As expected,
the ground state is fully polarised. We call the point $\mu_1 - \mu_2 = 4$
the first critical field.

Below the first critical field, i.e.\ if $\mu_1 - \mu_2 < 4$, the function
$\rep{e}{1}_1(x)$ possesses two symmetrically distributed real roots, and we
have $\rep{e}{n}_1(x) = \rep{e^+}{n}_1(x)$ only for $n = 2,3$. We can still
solve~\eqref{eq:nlielin} for the latter two functions and get
\begin{subequations}
  \begin{align}
    \rep{e}{1}_1(x) &= -\rep{V}{1}_{[2]}(x) + \frac{\mu_1 - \mu_2}{2} -
    \bigl[\rep{K}{1,1}_{[2]} \ast \rep{e^+}{1}_1\bigr](x)\msp,\\
    \rep{e}{2}_1(x) &= \rep{K}{1,1}_{[2]}(x) + \frac{1}{2} \sum_{j=1}^2 (\mu_j - \mu_3)
    - \bigl[\rep{V}{1}_{[2]} \ast \rep{e^+}{1}_1\bigr](x)\msp,\label{eq:critaux2}\\
    \rep{e}{3}_1(x) &= \mu_3 - \mu_4\msp.
  \end{align}
\end{subequations}
For the ground-state energy, we arrive at
\begin{equation}
  f_0 = 1 - 2 \ln 2 - \frac{\mu_1 + \mu_2}{2} - \bigl[\rep{V}{1}_{[2]} \ast \rep{e^+}{1}_1\bigr](x)\msp.
\end{equation}
This is exactly the $T = 0$ behaviour of the spin-$1/2$ Heisenberg
chain. Two states, $\alpha = 1$ and $2$, are present in the ground
state. Note that these equations are valid only above the second critical
field, i.e.\ as long as $\rep{e}{2}_1(x) \geq 0$ for all $x \in \set
R$. From~\eqref{eq:critaux2}, we find the restriction
\begin{equation}
  \sum_{j=1}^2 (\mu_j - \mu_3) \geq 4 \ln 2 + 2 \cdot \bigl[\rep{V}{1}_{[2]}
  \ast \rep{e^+}{1}_1\bigr](0)\msp,
\end{equation}
where the positive convolution term unfortunately still depends on the function
$\rep{e^+}{1}_1(x)$, which is not explicitly known. The
convolution term vanishes if $\mu_1 = \mu_2$.

Below the second critical field, the state $\alpha = 3$ also contributes to
the ground state. Both $\rep{e}{1}_1(x)$ and $\rep{e}{2}_1(x)$ possess two
real roots, and only $\rep{e}{3}_1(x) = \rep{e^+}{3}_1(x)$ remains
valid. Here, we recover the $T = 0$ behaviour of the $\alg{sl}(3)$-symmetric
US model,
\begin{subequations}
  \begin{align}
    \rep{e}{n}_1(x) &= -\rep{V}{n}_{[3]}(x) - \rep{c}{n}_1 - \sum_{m=1}^2
    \bigl[\rep{K}{n,m}_{[3]} \ast \rep{e^+}{m}_1\bigr](x) \qquad (n = 1,2)\msp,\\
    \rep{e}{3}_1(x) & = -W(x) + \frac{1}{3} \sum_{j=1}^3 (\mu_j - \mu_4) -
    \sum_{n=1}^{2} \bigl[\rep{V}{3-n}_{[3]} \ast \rep{e^+}{n}_1\bigr](x)\msp,\label{eq:critaux3}
  \end{align}
\end{subequations}
where $\rep{c}{n}_1$ are the constants of the $\alg{sl}(3)$ case defined
in~\eqref{eq:nliesl3const}. The ground-state energy can be calculated by use
of
\begin{equation}
  f_0 = 1 - \frac{\pi}{3 \sqrt{3}} - \ln 3 - \frac{1}{3} \sum_{j=1}^3 \mu_j -
  \sum_{n=1}^2 \bigl[\rep{V}{n}_{[3]} \ast \rep{e^+}{n}_1\bigr](0)\msp.
\end{equation}
These equations hold as long as we are above the third and last critical
field. We see from~\eqref{eq:critaux3} that the restriction is
\begin{equation}
  \sum_{j=1}^3 (\mu_j - \mu_4) \geq \pi \sqrt{3} - 3 \ln 3 + 3 \cdot \sum_{n=1}^{2}
  \bigl[\rep{V}{3-n}_{[3]} \ast \rep{e^+}{n}_1\bigr](0)\msp,
\end{equation}
where we again have no explicit expression for the positive convolution terms,
which vanish if $\mu_1 = \mu_2 = \mu_3$.

Finally, below the third and last critical field, all auxiliary functions possess two
symmetrically distributed real roots. Therefore,
equations~\eqref{eq:nlielin} and~\eqref{eq:nlielineigenv} have to be used
without further simplification. In this case, all four states contribute to
the ground state. We note that the equations~\eqref{eq:nlielin}
and~\eqref{eq:nlielineigenv} are equal to those one can get either from the
traditional TBA equations in the limit $T \to 0$ or directly from the BA
equations for the Hamiltonian~\cite{johannesson86b}.

\section{Numerical investigation}\label{sec:numerical}

The NLIE of the $\alg{sl}(4)$-symmetric Uimin-Sutherland model~\eqref{eq:nlie}
are of a type that allows for an efficient numerical solution by iteration and
use of the fast Fourier transform~(FFT) to calculate the convolutions. As
initial values, some discretised functions $\ln\rep{B}{n}_j(x)$ are taken. The
FFT is applied and the right-hand sides of the NLIE are computed in Fourier
space. After that, the functions $\ln\rep{b}{n}_j(x)$ are obtained using the
inverse FFT to eventually yield a new approximation for the functions
$\ln\rep{B}{n}_j(x)$. These steps have to be repeated, each time starting with
the previous approximation for $\ln\rep{B}{n}_j(x)$, until the numerical error
is small enough. Finally, the free energy is calculated
using~\eqref{eq:energy} and~\eqref{eq:nlielambda}.

From our NLIE it is also possible to directly calculate derivatives of the free
energy with respect to some parameter $p$. We just have to consider the
corresponding derivatives of the NLIE and of the expression for the
eigenvalue. For the first derivative, we exploit the relation
\begin{equation}
  \frac{\partial}{\partial p}\ln\rep{B}{n}_j(x) =
  \frac{\rep{b}{n}_j(x)}{\rep{B}{n}_j(x)} \cdot \frac{\partial}{\partial p}\ln{\rep{b}{n}_j(x)}
\end{equation}
and use an additional relation for the calculation of the second derivative,
\begin{equation}
  \frac{\partial^2}{\partial p^2}\ln\rep{B}{n}_j(x) =
  \frac{\rep{b}{n}_j(x)}{\rep{B}{n}_j(x)}
  \left\{\frac{1}{\rep{B}{n}_j(x)}\left(\frac{\partial}{\partial
        p}\ln\rep{b}{n}_j(x)\right)^2 + \frac{\partial^2}{\partial
      p^2}\ln\rep{b}{n}_j(x)\right\}\msp.
\end{equation}
In this case, the numerical calculation is done step by step. After solving
the unmodified NLIE to obtain $\ln\rep{B}{n}_j(x)$, its first and finally its
second derivative are solved in an analogous way. In each step, results from
the previous calculations are used.

In the following, we will provide some numerical results for a special application of
the one-dimensional $\alg{sl}(4)$-symmetric Uimin-Sutherland model. We consider
the Hamiltonian
\begin{equation}
  \opr H = \sum_{j=1}^L (2 \vect S_j \vect S_{j+1} + 1/2)(2 \vect \tau_j \vect
  \tau_{j+1} + 1/2) - \sum_{j=1}^L \left(g_S h S^z_j + g_\tau h \tau^z_j\right)\msp,
\end{equation}
which describes the $\alg{SU}(4)$-symmetric case of a $\alg{SU}(2) \times
\alg{SU}(2)$ spin-orbital model at the supersymmetric point. $\vect
S_j$ is a $\alg{SU}(2)$ spin-$1/2$ operator acting on the spins and
$\vect \tau_j$ is a $\alg{SU}(2)$ spin-$1/2$ operator acting on the orbital
pseudo-spin degrees of freedom. We have allowed for an external magnetic field
$h$, which couples to the spins and orbital pseudo-spins with Land\'e factors
$g_S$ and $g_\tau$, respectively. Clearly, the Hamiltonian
\label{eq:hspinorb} is equivalent to the $\alg{sl}(4)$-symmetric US
Hamiltonian~\eqref{eq:exthamilton} if we use the basis
\begin{align}
  \ket{1} &= \ket{\uparrow_S \uparrow_\tau}\msp, &
  \ket{2} &= \ket{\uparrow_S \downarrow_\tau}\msp, &
  \ket{3} &= \ket{\downarrow_S \uparrow_\tau}\msp, &
  \ket{4} &= \ket{\downarrow_S \downarrow_\tau}
\end{align}
and accordingly set
\begin{subequations}
  \begin{align}
    \mu_1 &= (g_S + g_\tau) h/2\msp, & \mu_2 &= (g_S - g_\tau) h/2\msp,\\
    \mu_3 &= -(g_S - g_\tau) h/2\msp, & \mu_4 &= -(g_S + g_\tau) h/2\msp.
  \end{align}
\end{subequations}

We are mainly interested in the entropy $S$, specific heat $C$,
magnetisation $M$ and magnetic susceptibility $\chi$, which are defined by
\begin{align}
  S &= -\frac{\partial f}{\partial T}\msp, &
  C &= -T \frac{\partial^2 f}{\partial T^2}\msp, &
  M &= -\frac{\partial f}{\partial h}\msp, &
  \chi &= -\frac{\partial^2 f}{\partial h^2}\msp.
\end{align}

\begin{figure}
  \centering
  \subfloat[Entropy vs.\ temperature.]{\includegraphics{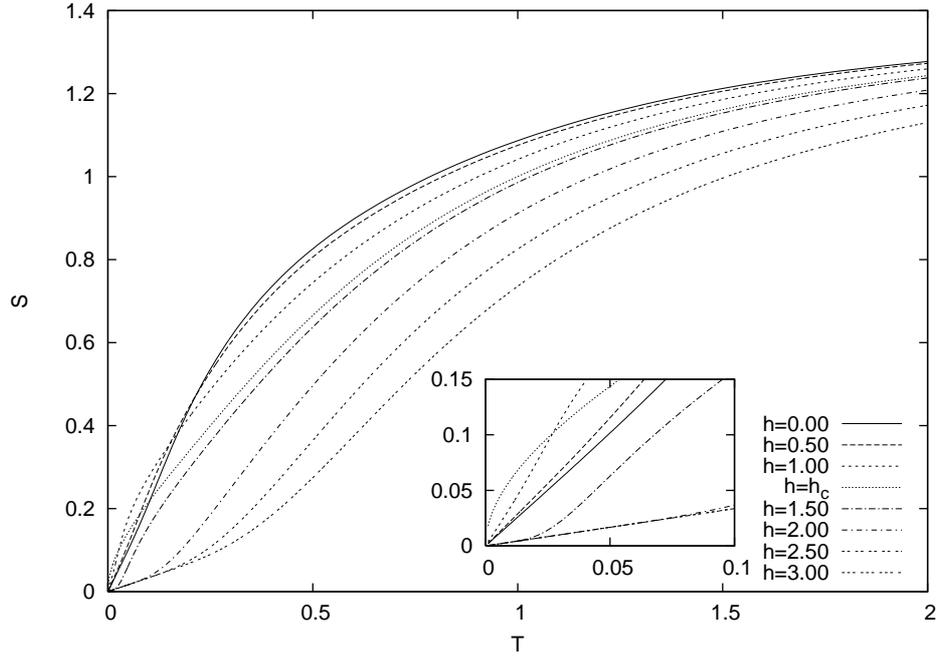}}\\
  \subfloat[Specific heat vs.\ temperature.]{\includegraphics{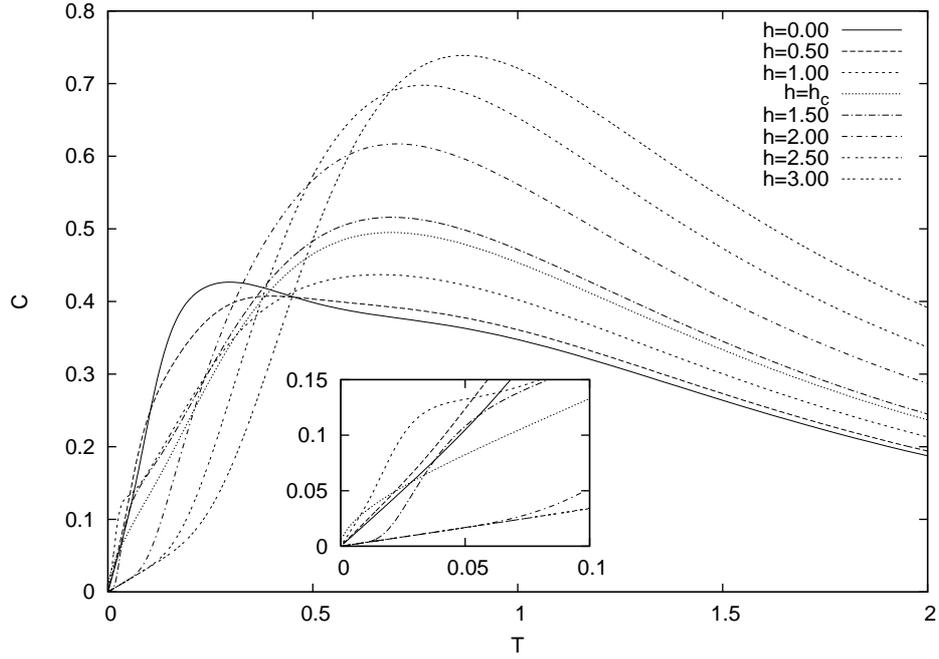}}
  \caption{Entropy and specific heat of the spin-orbital model at $g_S = 1$, $g_\tau =
    0$ for various magnetic fields. The insets show the low-temperature
    parts. The critical field is $h_c = 2 \ln 2 \approx 1.39$.}
  \label{fig:spinorb10a}
\end{figure}
\begin{figure}
  \centering
  \subfloat[Magnetisation vs.\ temperature.]{\includegraphics{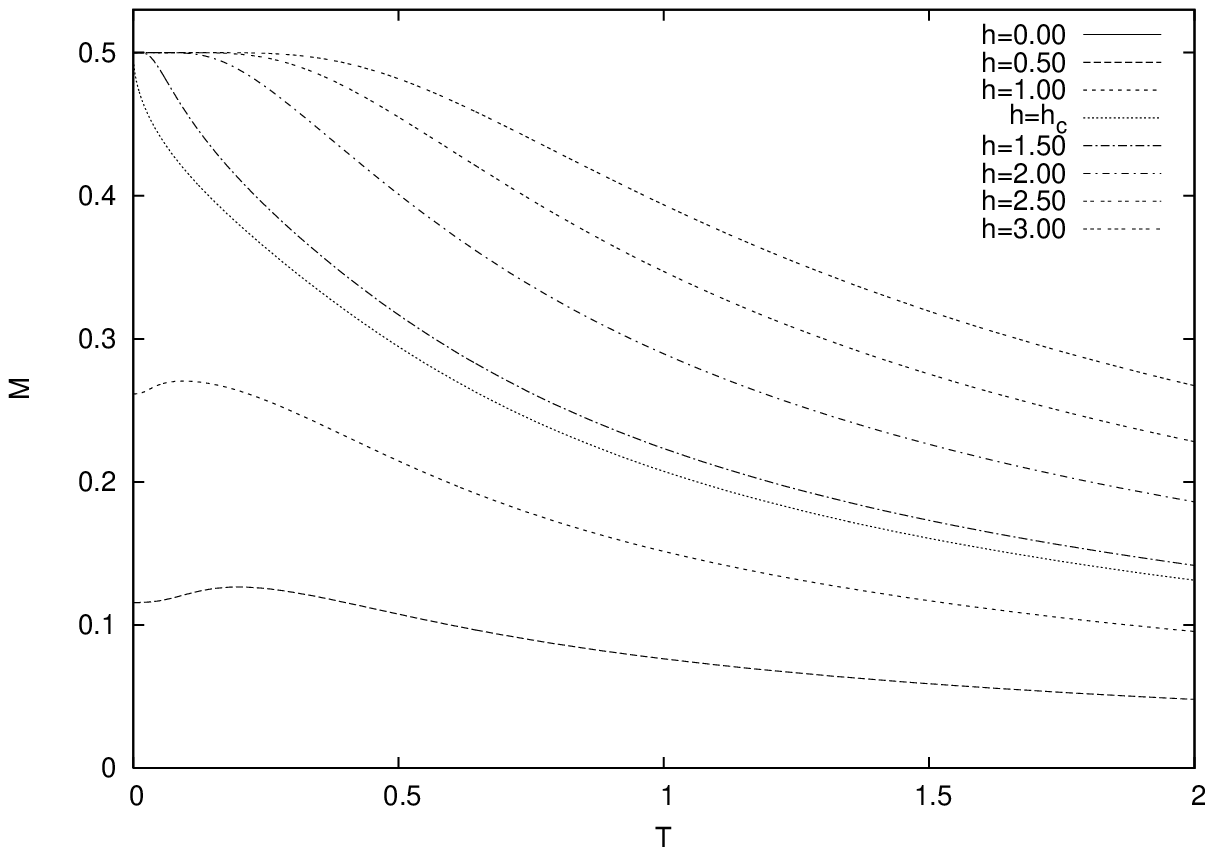}}\\
  \subfloat[Magnetic susceptibility vs.\ temperature.]{\includegraphics{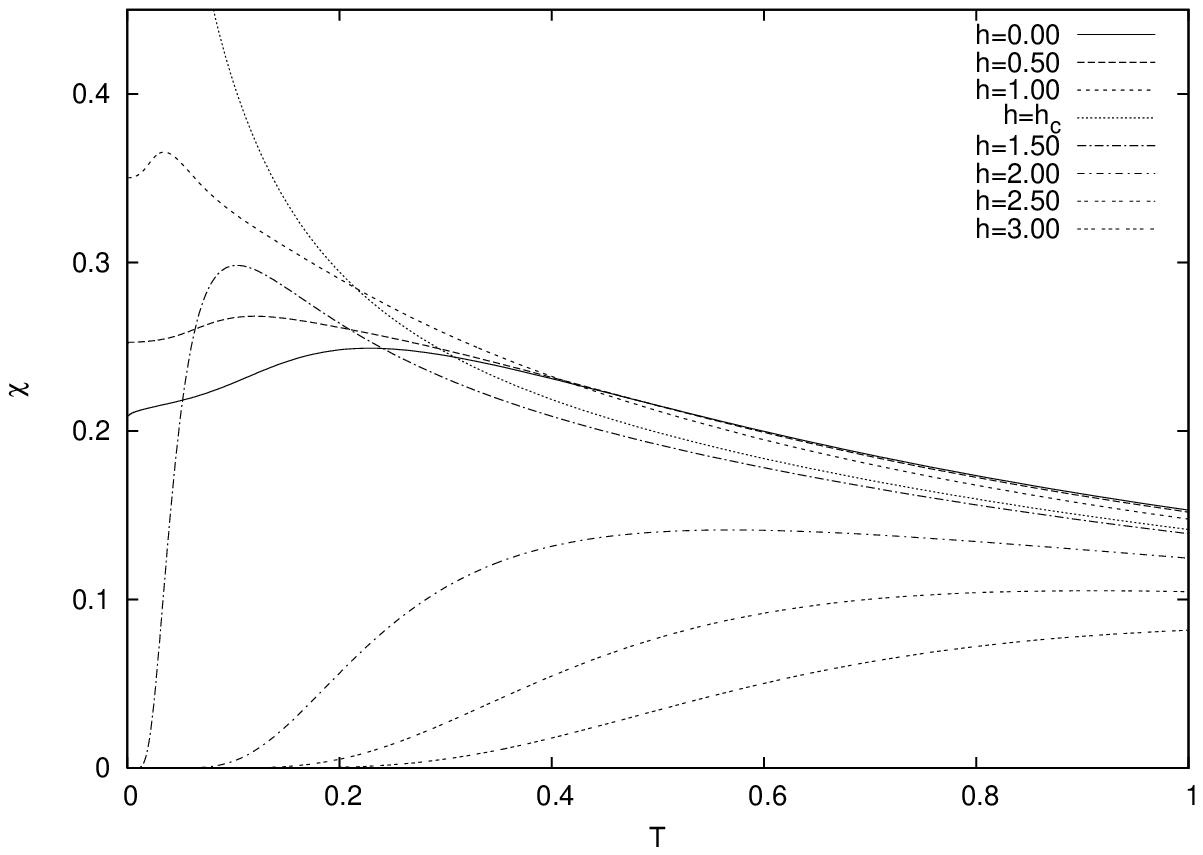}}
  \caption{Magnetisation and magnetic susceptibility of the spin-orbital model at $g_S = 1$, $g_\tau =
    0$ for various magnetic fields. The critical field is $h_c = 2 \ln 2
    \approx 1.39$.}
  \label{fig:spinorb10b}
\end{figure}
In Figure~\vref{fig:spinorb10a} and Figure~\vref{fig:spinorb10b}, results are shown for the case $g_S = 1$ and
$g_\tau = 0$, for which the magnetic field couples only to the spins.
In this case, we have $\mu_1 = \mu_2 = h/2$ and $\mu_3 = \mu_4 =
-h/2$. Therefore we know from our analytical investigation that there is
only one critical field exactly at $h_c = 2 \ln 2 \approx 1.39$. Below the critical
field, all four states contribute to the ground state. Above, the spins are
fully polarised and only the orbital degrees of freedom remain. Hence, the ground
state resembles that of the spin-$1/2$ Heisenberg chain. This phase transition
is clearly exposed by the numerical data. The low-temperature slopes both of
the entropy and the specific heat increase from $2$ at $h = 0$ to infinity at
$h = h_c$, whereas they are $1 / 3$ for $h > h_c$. Moreover, the magnetisation
data show the expected saturation behaviour for $h \geq h_c$. Notice also,
that the magnetic susceptibility diverges at the critical field. Below, the
value at $T = 0$ stays finite; above, it drops to zero.

\begin{figure}
  \centering
  \includegraphics{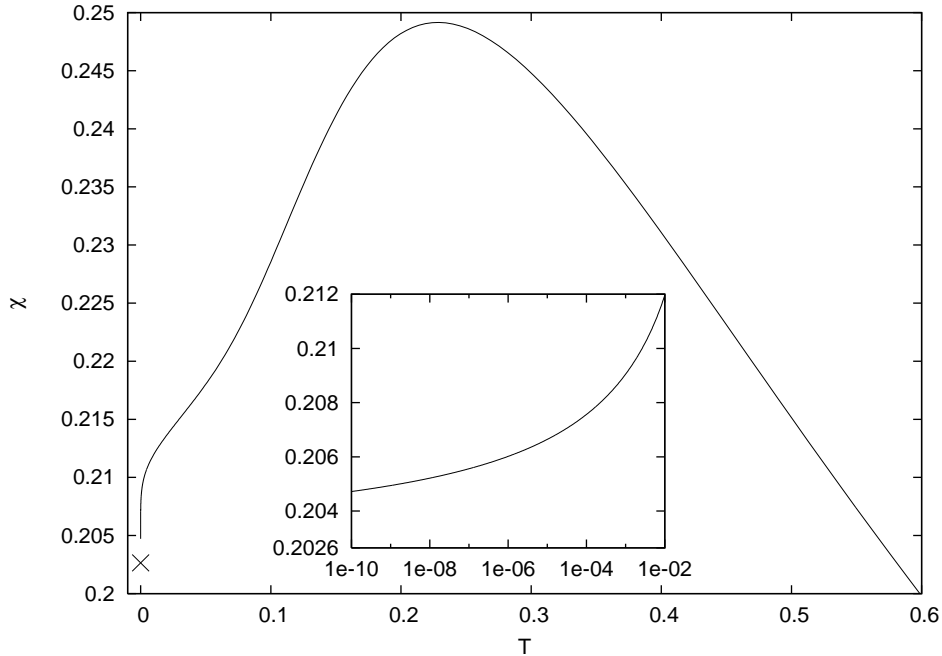}
  \caption{Magnetic susceptibility of the spin-orbital model at $h = 0$ for
    temperatures down to $T = 10^{-10}$. The cross denotes the ground-state
    value $\chi(0) = 2 / \pi^2 \approx 0.2026$. The inset shows the
    low-temperature part of the susceptibility using a logarithmic scale.}
  \label{fig:su4_spinorb_logcorr}
\end{figure}
The magnetic susceptibility at $h = 0$ is particularly interesting as it
is expected to show a characteristic singular behaviour at $T = 0$ due to logarithmic
corrections. Indeed, this is confirmed by our results for the low-temperature susceptibility, see
Figure~\vref{fig:su4_spinorb_logcorr}. Even for the lowest plotted
temperature, $T = 10^{-10}$, the susceptibility is still well above the
ground-state value $\chi(0) = 2 / \pi^2$.
For the spin-$1/2$ Heisenberg model, these corrections have already been treated in
detail~\cite{eggert94,kluemper98,lukyanov98,kluemper00}; similar results
for $\alg{sl}(q)$-symmetric US models are known~\cite{schlottmann92,fujii99}. 

\begin{figure}
  \centering
  \subfloat[Entropy vs.\ temperature.]{\includegraphics{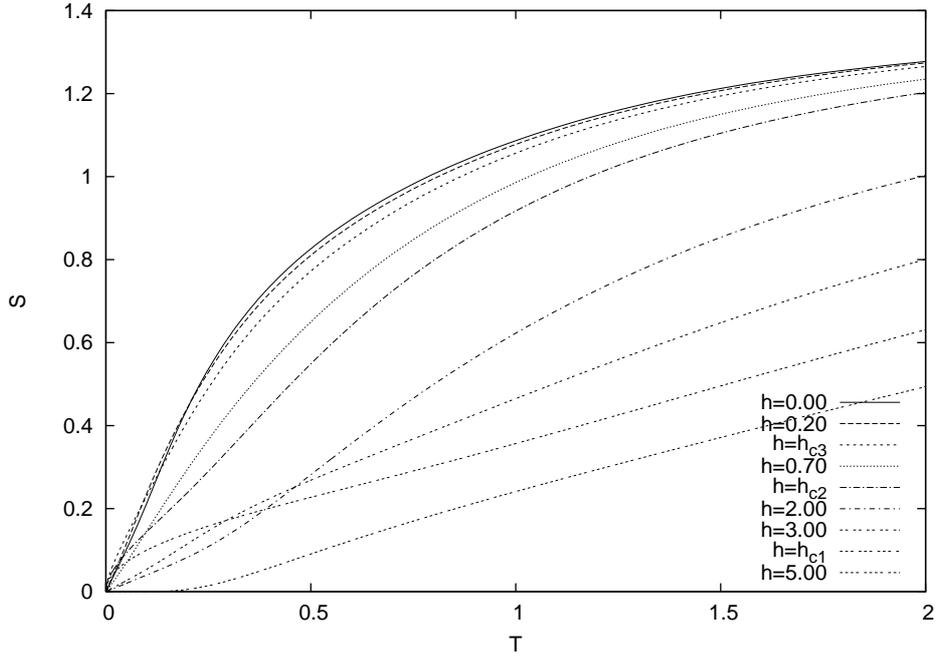}}\\
  \subfloat[Specific heat vs.\ temperature.]{\includegraphics{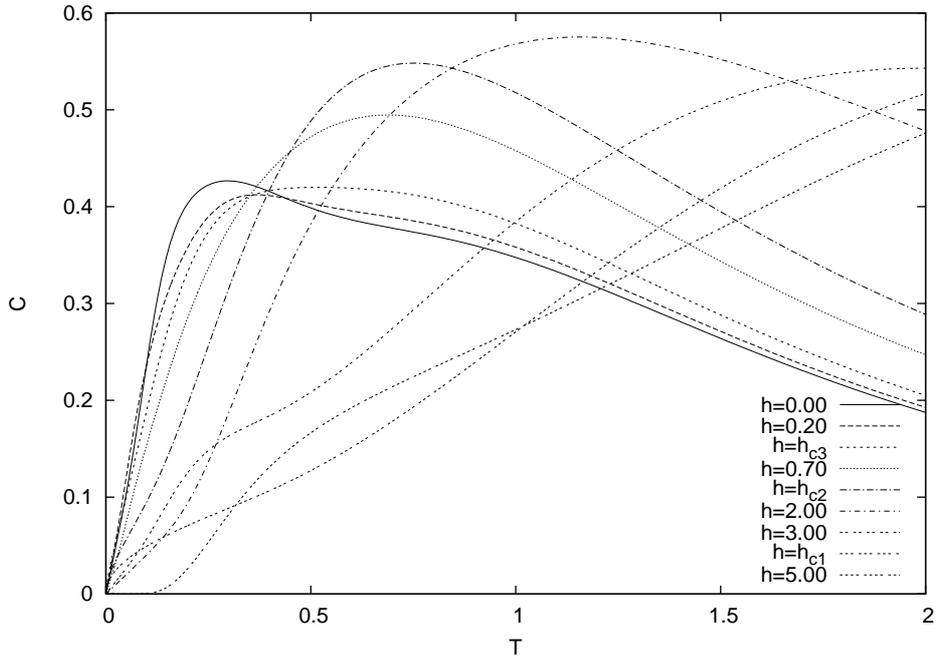}}
  \caption{Entropy and specific heat of the spin-orbital model at $g_S = 1$, $g_\tau =
    2$ for various magnetic fields. The critical fields are $h_{c1} = 4$, $h_{c2} \approx 0.941$ and
    $h_{c3} \approx 0.370$. Note the numbering of the critical fields as
    discussed \vpageref{txt:hcnumbering}.}
  \label{fig:spinorb12a}
\end{figure}
\begin{figure}
  \centering
  \subfloat[Magnetisation vs.\ temperature.]{\includegraphics{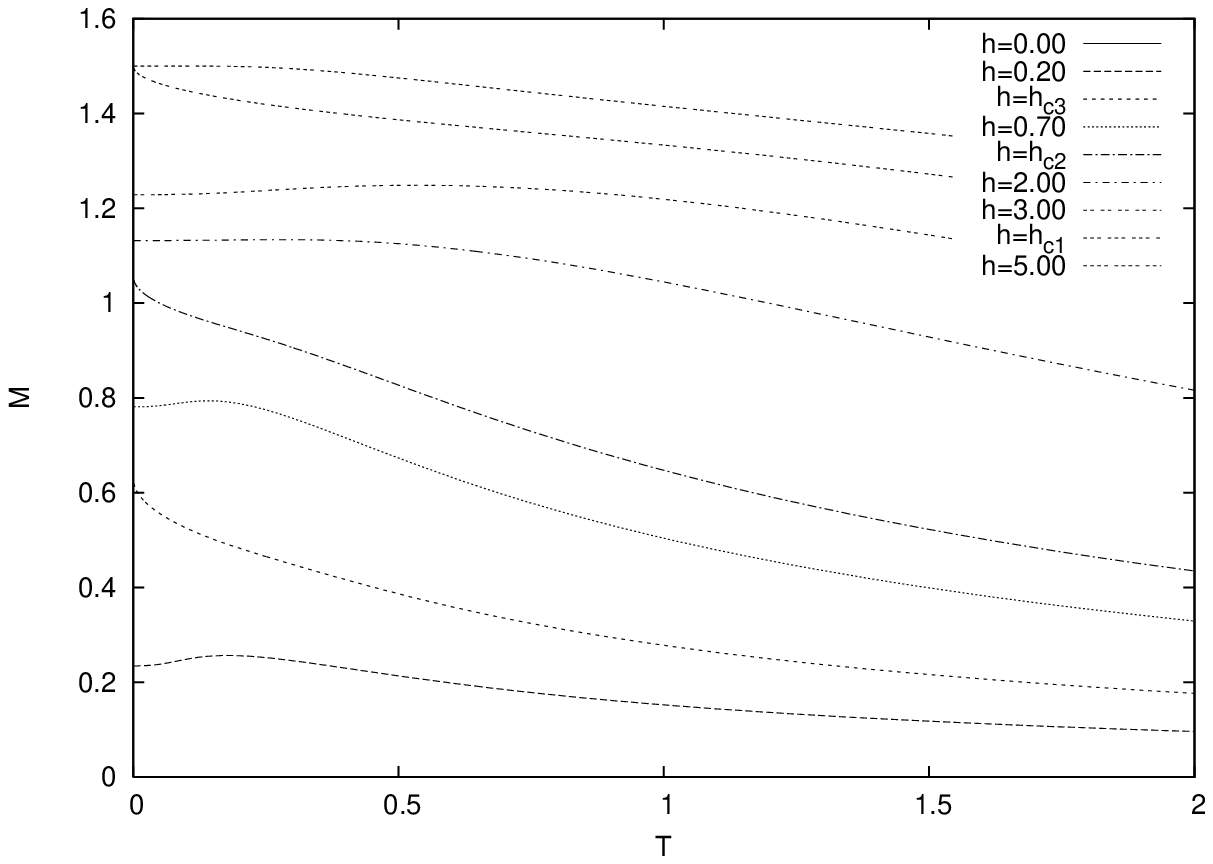}}\\
  \subfloat[Magnetic susceptibility vs.\ temperature.]{\includegraphics{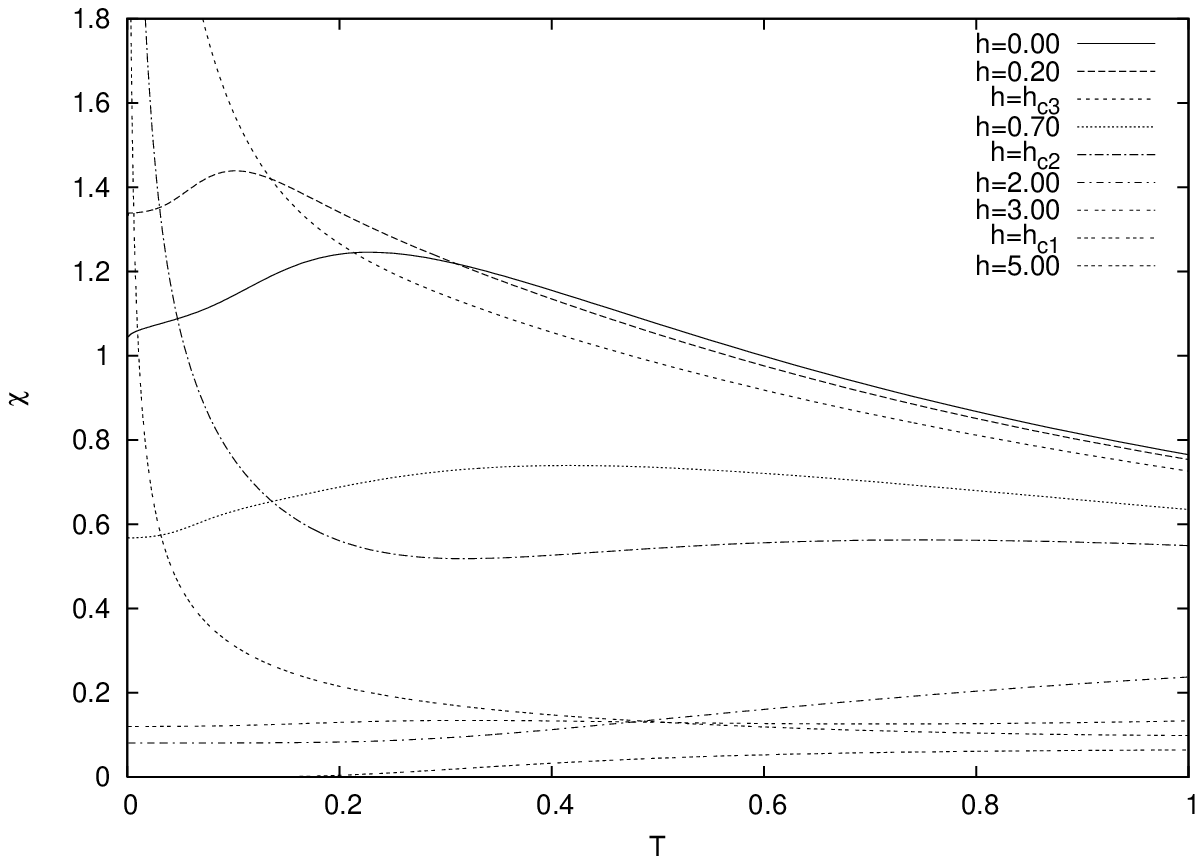}}
  \caption{Magnetisation and magnetic susceptibility of the spin-orbital model at $g_S = 1$, $g_\tau =
    2$ for various magnetic fields. The critical fields are $h_{c1} = 4$, $h_{c2} \approx 0.941$ and
    $h_{c3} \approx 0.370$. Note the numbering of the critical fields as
    discussed \vpageref{txt:hcnumbering}.}
  \label{fig:spinorb12b}
\end{figure}
We also consider the spin orbital-model at $g_S = 1$ and
$g_\tau = 2$, which corresponds to the spin-$3/2$ interpretation. Here we have
all three possible types of phase transitions. Numerical results for this case
showing the rich resulting structure are plotted in
Figure~\vref{fig:spinorb12a} and Figure~\vref{fig:spinorb12b}.
Again, we observe that the low-temperature susceptibility shows characteristic
singular behaviour at $h = 0$ and diverges at the critical fields. The
highest critical field is at $h_{c1} = 4$, the other two have to be calculated
numerically as only the lower bounds $h_{c2} > 4 \ln(2) / 3 \approx 0.924$ and
$h_{c3} > \pi / (2 \sqrt{3}) - \ln(3) / 2 \approx 0.358$ are known
explicitly. We find the remaining critical fields to be $h_{c2} \approx 0.941$
and $h_{c3} \approx 0.370$.
\phantomsection\label{txt:hcnumbering}Note the numbering of the critical
fields, where at each field $h_{cj}$ the number of involved degrees of freedom
changes from $j$ to $j + 1$. The advantage of this scheme is that the
critical fields $h_{c1}$ and $h_{c2}$ also appear in the spin-$1$
interpretation of the $\alg{sl}(3)$-symmetric US model~\cite{fujii99}, while
only $h_{c1}$ remains in the spin-$1/2$ Heisenberg model. 

\begin{figure}
  \centering
  \includegraphics{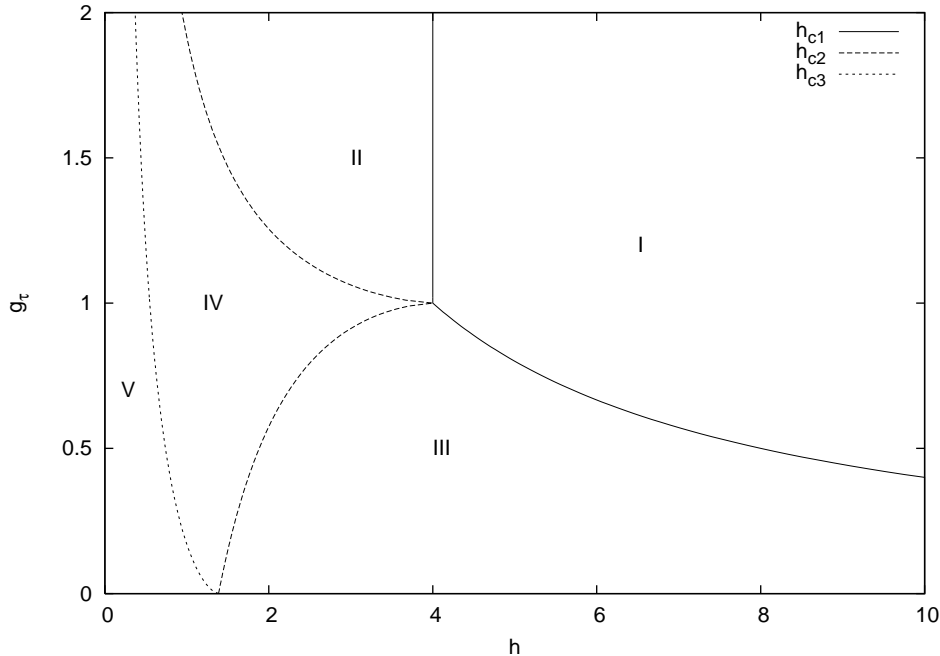}
  \caption{Phase diagram of the spin-orbital model for magnetic field $h$
  and Land\'e factor $g_\tau$, where $g_S = 1$ is held constant. There exist
  five different phases (I--V), for details see the text \vpageref{txt:su4_phase}.}
  \label{fig:su4_phase}
\end{figure}
Figure~\vref{fig:su4_phase} shows the complete phase diagram
of the spin-orbital model depending on the magnetic field $h$ and the orbital
Land\'e factor $g_\tau$, while the Land\'e factor of the spins is set to $g_S
= 1$.
\phantomsection\label{txt:su4_phase}Obviously, there exist five different
phases. Above $h_{c1}$~(I), all spins and orbitals are fully
polarised. Between $h_{c1}$ and $h_{c2}$, we have to distinguish the regions
$g_\tau > 1$~(II) and $g_\tau < 1$~(III). In the first region, the orbitals
are fully polarised, while the spins are only partially aligned. In the latter
case, it is the other way around. For $g_\tau = 1$, we have a direct
transition from phase I to phase IV, because $h_{c1} = h_{c2}$. For a magnetic
field below $h_{c2}$, but above $h_{c3}$~(IV), the spins and orbitals are both
partially polarised, while the state $\ket{\downarrow_S \downarrow_\tau}$ is
still completely suppressed. For $h < h_{c3}$~(V), all possible spin
configurations contribute to the ground state. Notice that $h_{c1}$ tends to
infinity for $g_\tau \to 0$. As we have seen before, only one phase transition
survives for $g_\tau = 0$, where the magnetic field couples only to the
spins. The phase diagram presented here is qualitatively in perfect agreement
with~\cite{gu02}, where a finite system of 200 sites has been used for the
calculation.

\section{Summary and outlook}\label{sec:summary}

We presented a set of suitable auxiliary functions that allowed us to derive a finite
set of NLIE for the thermodynamics of the $\alg{sl}(4)$-symmetric
one-dimensional US model. These NLIE are well posed for an efficient numerical treatment
using the fast Fourier transform as they are of
convolution type. Moreover, they are valid for the complete temperature
range and arbitrary chemical potentials. We have provided analytical results for
several limiting cases of our NLIE. Thus, we recovered the
previously known NLIE of the $\alg{sl}(3)$ case and the linearised integral
equations for $T = 0$. Both provide further support for the validity of the
NLIE. Using the latter limit, we were able to derive all critical fields. We
also gave some numerical results for the $\alg{SU}(4)$ spin-orbital model as an
example for the application of the $\alg{sl}(4)$-symmetric US model, achieving
high accuracy even at low temperatures.

This is the first time that NLIE of this type were derived for the
$\alg{sl}(q)$-symmetric US model with $q > 3$. Unfortunately, up to now there is
no general scheme to construct the needed auxiliary functions for the case $q
\geq 5$. The functions presented here were basically found by trial and
error. Even at the level of the NLIE, it proves difficult to generalise the
structure of the somewhat complicated kernel matrix. Once the NLIE are
known for some number of states $q$, there is a straightforward way to extract the NLIE
of all $\alg{sl}(n)$-symmetric cases with $n < q$ by freezing out one or more
of the states. Going the opposite way fixes some of the structure, but does
not provide enough information to deduce the entries of the kernel matrix
lying on the anti-diagonal.  Nevertheless, we hope to overcome these
difficulties and to generalise the approach to cover the whole class of
$\alg{sl}(q)$-symmetric models in the near future. For that, we expect that
the number of required auxiliary functions will be equal to the sum of the
dimensions of all fundamental representations of $\alg{sl}(q)$, i.e.\ $2^q -
2$. We note that a further generalisation of this type of NLIE from the
fundamental representation to higher rank representations~\cite{andrei84}
should be quite straightforward. To achieve this, one has to consider slightly
modified auxiliary functions, which can be used to truncate the TBA equations
at an arbitrary step, in analogy to~\cite{suzuki99}.

Another open question is the connection between the kernel matrix and the
complete $S$-matrix of elementary excitations~\cite{doikou98}. In the $T = 0$
case, all kernel functions can be obtained from corresponding $S$-matrix
entries~\cite{kulish81}, the generalisation to $T > 0$ is unknown.

The treatment of graded models is a further direction of generalisation. Up to
now, only the $\alg{sl}(2|1)$-symmetric case has been treated this
way~\cite{juettner97a,juettner97b}. Although supersymmetric models are
generally more difficult, it turns out that, at least for the $\alg{sl}(2|1)$
case, the corresponding NLIE are even simpler than the NLIE of the
$\alg{sl}(3)$ case as only three auxiliary functions are needed. The NLIE for
the graded $q = 4$ models, e.g.\ the $\alg{sl}(2|2)$ case~\cite{essler92a},
also seem to be simpler than the NLIE presented here. We hope to report soon
on details concerning this model.

\section*{Acknowledgments}

The authors like to acknowledge support by the research program of the
Graduiertenkolleg 1052 funded by the Deutsche
Forschungsgemeinschaft.

\appendix

\section{Derivation of the NLIE}\label{app:nliederiv}

The auxiliary functions, as defined in~\eqref{eq:aux1}--\eqref{eq:aux3}, are
rational functions. Moreover, we have numerically checked that they are
analytic, non-zero and have constant asymptotics~(ANZC) in a strip $-1/2
\lesssim \Im(x) \lesssim 1/2$, which includes the real axis. Therefore, we are
allowed to apply a Fourier transform to the logarithmic derivative of all
auxiliary functions,
\begin{equation}\label{eq:logft}
  \widehat f(k) = \int_{-\infty}^{\infty} \frac{\dif}{\dif x}\left[\ln f(x)\right]
    \ex{-\im k x}\,\frac{\dif x}{2 \pi}\msp.
\end{equation}
In the cases $k < 0$ and $k > 0$, we can close the integration path to a contour
above and below the real axis, respectively. Hence, it is important to
know in which of these regions the first-order poles of the logarithmic derivatives are
located and how they can be further classified.

Let us first introduce some notation. We define the constants
\begin{subequations}
  \begin{align}
    \rep{a}{1} &= \ex{\beta \mu_1} + \ex{\beta \mu_2} + \ex{\beta \mu_3} +
    \ex{\beta \mu_4}\msp,\\
    \begin{split}
      \rep{a}{2} &= \ex{\beta (\mu_1 + \mu_2)} + \ex{\beta (\mu_1 + \mu_3)} +
      \ex{\beta (\mu_1 + \mu_4)}\\
      &\qquad + \ex{\beta (\mu_2 + \mu_3)} + \ex{\beta (\mu_2
        + \mu_4)} + \ex{\beta (\mu_3 + \mu_4)}\msp,
    \end{split}\\
    \rep{a}{3} &= \ex{\beta (\mu_1 + \mu_2 + \mu_3)} + \ex{\beta (\mu_1 +
      \mu_2 + \mu_4)} + \ex{\beta (\mu_1 + \mu_3 + \mu_4)} + \ex{\beta (\mu_2
      + \mu_3 + \mu_4)}\msp,\\
    \varphi_1 &= \ex{\beta \mu_1} + \ex{\beta \mu_2}\msp,\qquad
    \varphi_2 = \ex{\beta \mu_2} + \ex{\beta \mu_3}\msp,\qquad
    \varphi_3 = \ex{\beta \mu_3} + \ex{\beta \mu_4}\msp,\\
    \rep{\chi}{1}_1 &= \ex{\beta \mu_1} + \ex{\beta \mu_2} + \ex{\beta
      \mu_3}\msp,\qquad
    \rep{\chi}{1}_2 = \ex{\beta \mu_2} + \ex{\beta \mu_3} + \ex{\beta
      \mu_4}\msp,\qquad\\
    \rep{\chi}{2}_1 &= \ex{\beta (\mu_1 + \mu_2)} + \ex{\beta (\mu_1 + \mu_3)} +
    \ex{\beta (\mu_1 + \mu_4)} + \ex{\beta (\mu_2 + \mu_3)} + \ex{\beta (\mu_2
      + \mu_4)}\msp,\\
    \rep{\chi}{2}_2 &= \ex{\beta (\mu_1 + \mu_3)} + \ex{\beta (\mu_1 + \mu_4)} +
    \ex{\beta (\mu_2 + \mu_3)} + \ex{\beta (\mu_2 + \mu_4)} + \ex{\beta (\mu_3
      + \mu_4)}\msp,\\
    \rep{\chi}{3}_1 &= \ex{\beta (\mu_1 + \mu_2)} + \ex{\beta (\mu_1 + \mu_3)} +
    \ex{\beta (\mu_2 + \mu_3)}\msp,\\
    \rep{\chi}{3}_2 &= \ex{\beta (\mu_2 + \mu_3)} + \ex{\beta (\mu_2 + \mu_4)} +
    \ex{\beta (\mu_3 + \mu_4)}\msp.
  \end{align}
\end{subequations}
For the eigenvalues~\eqref{eq:fundeigenv}, we find the factorisation
\begin{subequations}
  \begin{align}
    \rep{\Lambda}{1}(x) &= \rep{\widetilde\Lambda}{1}(x) \cdot \rep{a}{1}\msp,\\
    \rep{\Lambda}{2}(x) &= \phi_-(x + \im / 2) \phi_+(x - \im / 2) \rep{\widetilde\Lambda}{2}(x) \cdot \rep{a}{2}\msp,\\
    \rep{\Lambda}{3}(x) &= \phi_-(v) \phi_-(v + \im) \phi_+(v - \im) \phi_+(v)
    \rep{\widetilde\Lambda}{3}(x) \cdot \rep{a}{3}\msp,
  \end{align}
\end{subequations}
where all $\rep{\widetilde\Lambda}{n}(x)$ are polynomials of degree $N$ with
the highest coefficients being one. For the other terms in the auxiliary
functions, which are generated by sums of certain Young tableaux,
we also find that several of their potential poles vanish due to the BA
equations~\eqref{eq:baeq}. We can therefore write
\begin{subequations}
  \begin{align}
    \left.\young(1) + \young(2)\right|_{v=x} &= \frac{\phi_+(x)
      q_1^{(h)}(x)}{q_2(x)} \cdot \varphi_1\msp,\\
    \left.\young(2) + \young(3)\right|_{v=x} &= \frac{\phi_-(x) \phi_+(x)
      q_2^{(h)}(x)}{q_1(x) q_3(x)} \cdot \varphi_2\msp,\\
    \left.\young(3) + \young(4)\right|_{v=x} &= \frac{\phi_-(x)
      q_3^{(h)}(x)}{q_2(x)} \cdot \varphi_3\msp,\\
    \left.\young(1) + \young(2) + \young(3)\right|_{v=x} &= \frac{\phi_+(x)
      \rep{X}{1}_1(x)}{q_3(x)} \cdot \rep{\chi}{1}_1\msp,\\
    \left.\young(2) + \young(3) + \young(4)\right|_{v=x} &= \frac{\phi_-(x)
      \rep{X}{1}_2(x)}{q_1(x)} \cdot \rep{\chi}{1}_2\msp,\\
    \left.\young(1,2) + \young(1,3) + \young(1,4) + \young(2,3) +
      \young(2,4)\right|_{v=x} &= \frac{\phi_-(x+\frac{\im}{2})
      \phi_+(x-\frac{\im}{2}) \rep{X}{2}_1(x)}{q_2(x-\frac{\im}{2})} \cdot
    \rep{\chi}{2}_1\msp,\\
    \left.\young(1,3) + \young(1,4) + \young(2,3) + \young(2,4) +
      \young(3,4)\right|_{v=x} &= \frac{\phi_-(x+\frac{\im}{2})
      \phi_+(x-\frac{\im}{2}) \rep{X}{2}_2(x)}{q_2(x+\frac{\im}{2})} \cdot
    \rep{\chi}{2}_2\msp,\\
    \left.\young(1,2) + \young(1,3) + \young(2,3)\right|_{v=x} &=
    \frac{\phi_-(x+\frac{\im}{2}) \phi_+(x-\frac{\im}{2})
      \phi_+(x+\frac{\im}{2}) \rep{X}{3}_1(x)}{q_3(x+\frac{\im}{2})} \cdot
    \rep{\chi}{3}_1\msp,\\
    \left.\young(2,3) + \young(2,4) + \young(3,4)\right|_{v=x} &=
    \frac{\phi_-(x+\frac{\im}{2}) \phi_+(x-\frac{\im}{2})
      \phi_-(x-\frac{\im}{2}) \rep{X}{3}_2(x)}{q_1(x-\frac{\im}{2})} \cdot
    \rep{\chi}{3}_2\msp.
  \end{align}
\end{subequations}
All functions $q_j^{(h)}(x)$ and $\rep{X}{n}_j(x)$ are polynomials of degree
$N$, with the exception of $\rep{X}{2}_1(x)$
and $\rep{X}{2}_2(x)$ being of degree $3N/2$, and with a highest coefficient
of one. We note that the roots of the functions $q_j^{(h)}(x)$ provide
additional solutions to the corresponding BA equations~\eqref{eq:baeq} and are
called the hole-type solutions.

The roots of the polynomials can be obtained from numerical solutions
of the BA equations at finite $N$. We find that, in the complex plane, all
roots are located on slightly curved lines close to horizontal axes at certain
heights. Each curve contains $N/2$ many roots. The corresponding height values
for each polynomial are given in the following table:
\begin{subequations}
  \begin{align}
    q_j(x)&\text{: }0 &
    \rep{\widetilde\Lambda}{j}(x)&\text{: $\pm(j+1)/2$}\\
    q_j^{(h)}(x)&\text{: $\pm 1$} &
    \rep{X}{1}_j(x)&\text{: $\pm 1$}\\
    \rep{X}{2}_1(x)&\text{: $+1/2$, $\pm 3/2$} &
    \rep{X}{2}_2(x)&\text{: $-1/2$, $\pm 3/2$}\\
    \rep{X}{3}_j(x)&\text{: $\pm 3/2$}
  \end{align}
\end{subequations}
We like to stress that the deviations from these axes remain small
even for large $N$.

Next, we write the auxiliary functions $\rep{b}{n}_j(x)$ in
a factorised form, from which one can easily read off the locations of all
roots and poles,
\begin{subequations}\label{eq:auxfact1}
  \begin{align}
    \rep{b}{1}_1(x) &= \frac{\phi_-(x-\frac{\im}{2}) \phi_+(x+\frac{\im}{2})
      q_1(x+\frac{3}{2}\im)}{\phi_-(x+\frac{\im}{2})
      \rep{X}{1}_2(x+\frac{\im}{2})} \cdot \frac{\ex{\beta\mu_1}}{\rep{\chi}{1}_2}\msp,\\
    \rep{b}{1}_2(x) &= \frac{\phi_-(x-\frac{\im}{2}) \phi_+(x+\frac{\im}{2})
      q_2(x+\frac{3}{2}\im)\rep{X}{3}_2(x)}{q_1(x+\frac{\im}{2}) q_3^{(h)}(x+\frac{\im}{2})
      \rep{\widetilde\Lambda}{2}(x)} \cdot \frac{\ex{\beta\mu_2}
      \rep{\chi}{3}_2}{\varphi_3 \rep{a}{2}}\msp,\\
    \rep{b}{1}_3(x) &= \frac{\phi_-(x-\frac{\im}{2}) \phi_+(x+\frac{\im}{2}) q_2(x-\frac{3}{2}\im)
      q_3(x+\frac{3}{2}\im)}{q_3(x-\frac{\im}{2}) \rep{X}{2}_2(x)} \cdot \frac{\ex{2\beta\mu_3}}{\rep{\chi}{2}_2}\msp,\\
    \rep{b}{1}_4(x) &= \frac{\phi_-(x-\frac{\im}{2}) \phi_+(x+\frac{\im}{2})
      q_3(x-\frac{3}{2}\im)}{\phi_+(x-\frac{\im}{2})
      \rep{X}{1}_1(x-\frac{\im}{2})} \cdot \frac{\ex{\beta\mu_4}}{\rep{\chi}{1}_1}\msp,\\
    \rep{b}{2}_1(x) &= \frac{\phi_-(x-\im) \phi_+(x+\im)
      q_2(x+2\im)}{\rep{X}{2}_2(x+\frac{\im}{2})} \cdot \frac{\ex{\beta(\mu_1+\mu_2)}}{\rep{\chi}{2}_2}\msp,\\
    \rep{b}{2}_2(x) &= \frac{\phi_-(x-\im) \phi_+(x+\im) q_1(x+\im) q_2(x-\im)
      q_3(x+2\im)}{q_2(x+\im) \rep{X}{1}_1(x) \rep{X}{3}_2(x + \frac{\im}{2})}
    \cdot \frac{\ex{\beta(\mu_1+2\mu_3)}}{\rep{\chi}{1}_1 \rep{\chi}{3}_2}\msp,\\
    \rep{b}{2}_3(x) &= \frac{\phi_-(x-\im) \phi_+(x+\im) q_1(x+\im)
      q_3(x-\im)}{q_2^{(h)}(x) \rep{\widetilde\Lambda}{1}(x)}
    \cdot \frac{\ex{\beta(\mu_1+\mu_4)}}{\varphi_2 \rep{a}{1}}\msp,\\
    \rep{b}{2}_4(x) &= \frac{\phi_-(x-\im) \phi_+(x+\im) q_1(x-2\im)
      q_3(x+2\im)}{q_2^{(h)}(x) \rep{\widetilde\Lambda}{3}(x)}
    \cdot \frac{\ex{2\beta(\mu_2 + \mu_3)}}{\varphi_2 \rep{a}{3}}\msp,\\
    \rep{b}{2}_5(x) &= \frac{\phi_-(x-\im) \phi_+(x+\im) q_1(x-2\im)
      q_2(x+\im) q_3(x-\im)}{q_2(x-\im) \rep{X}{1}_2(x) \rep{X}{3}_1(x-\frac{\im}{2})}
    \cdot \frac{\ex{\beta(2\mu_2+\mu_4)}}{\rep{\chi}{1}_2 \rep{\chi}{3}_1}\msp,\\
    \rep{b}{2}_6(x) &= \frac{\phi_-(x-\im) \phi_+(x+\im)
      q_2(x-2\im)}{\rep{X}{2}_1(x-\frac{\im}{2})} \cdot \frac{\ex{\beta(\mu_3+\mu_4)}}{\rep{\chi}{2}_1}\msp,\\
    \rep{b}{3}_1(x) &= \frac{\phi_-(x-\frac{3}{2}\im) \phi_+(x+\frac{3}{2}\im)
      q_3(x+\frac{5}{2}\im)}{\phi_+(x+\frac{5}{2}\im) \rep{X}{3}_1(x)}
    \cdot \frac{\ex{\beta(\mu_1+\mu_2+\mu_3)}}{\ex{\beta\mu_4} \rep{\chi}{3}_1}\msp,\\
    \rep{b}{3}_2(x) &= \frac{\phi_-(x-\frac{3}{2}\im) \phi_+(x+\frac{3}{2}\im) q_2(x+\frac{3}{2}\im)
      q_3(x-\frac{\im}{2})}{q_3(x+\frac{3}{2}\im) \rep{X}{2}_1(x)} \cdot \frac{\ex{\beta(\mu_1 + \mu_2
      + \mu_4)}}{\ex{\beta\mu_3} \rep{\chi}{2}_1}\msp,\\
    \rep{b}{3}_3(x) &=\frac{\phi_-(x-\frac{3}{2}\im) \phi_+(x+\frac{3}{2}\im) q_2(x-\frac{3}{2}\im)
      \rep{X}{1}_2(x+\frac{\im}{2})}{q_1(x-\frac{3}{2}\im) q_3^{(h)}(x+\frac{\im}{2})
      \rep{\widetilde\Lambda}{2}(x)}
    \cdot \frac{\ex{\beta(\mu_1+\mu_3+\mu_4)} \rep{\chi}{1}_2}{\ex{\beta\mu_2}
      \varphi_3 \rep{a}{2}}\msp,\\
    \rep{b}{3}_4(x) &=\frac{\phi_-(x-\frac{3}{2}\im) \phi_+(x+\frac{3}{2}\im)
      q_1(x-\frac{5}{2}\im)}{\phi_-(x-\frac{5}{2}\im) \rep{X}{3}_2(x)} \cdot
    \frac{\ex{\beta(\mu_2+\mu_3+\mu_4)}}{\ex{\beta\mu_1} \rep{\chi}{3}_2}\msp.
  \end{align}
\end{subequations}
The uppercase auxiliary functions $\rep{B}{n}_j(x)$ factorise in an analogous way,
\begin{subequations}\label{eq:auxfact2}
  \begin{align}
    \rep{B}{1}_1(x) &= \frac{q_1(x+\frac{\im}{2}) \rep{\widetilde\Lambda}{1}(x+\frac{\im}{2})}{\phi_-(x+\frac{\im}{2})
      \rep{X}{1}_2(x+\frac{\im}{2})} \cdot \frac{\rep{a}{1}}{\rep{\chi}{1}_2}\msp,\\
    \rep{B}{1}_2(x) &= \frac{\rep{X}{1}_2(x+\frac{\im}{2}) \rep{X}{2}_2(x)}{q_1(x+\frac{\im}{2}) q_3^{(h)}(x+\frac{\im}{2})
      \rep{\widetilde\Lambda}{2}(x)} \cdot \frac{\rep{\chi}{1}_2
      \rep{\chi}{2}_2}{\varphi_3 \rep{a}{2}}\msp,\\
    \rep{B}{1}_3(x) &= \frac{q_3^{(h)}(x+\frac{\im}{2})
      \rep{X}{1}_1(x-\frac{\im}{2})}{q_3(x-\frac{\im}{2}) \rep{X}{2}_2(x)} \cdot
    \frac{\varphi_3 \rep{\chi}{1}_1}{\rep{\chi}{2}_2}\msp,\\
    \rep{B}{1}_4(x) &= \frac{q_3(x-\frac{\im}{2}) \rep{\widetilde\Lambda}{1}(x-\frac{\im}{2})}{\phi_+(x-\frac{\im}{2})
      \rep{X}{1}_1(x-\frac{\im}{2})} \cdot \frac{\rep{a}{1}}{\rep{\chi}{1}_1}\msp,\\
    \rep{B}{2}_1(x) &= \frac{q_2(x+\im)
      \rep{\widetilde\Lambda}{2}(x+\frac{\im}{2})}{\rep{X}{2}_2(x+\frac{\im}{2})}
    \cdot \frac{\rep{a}{2}}{\rep{\chi}{2}_2}\msp,\\
    \rep{B}{2}_2(x) &= \frac{q_2^{(h)}(x)
      \rep{X}{2}_2(x+\frac{\im}{2})}{q_2(x+\im) \rep{X}{1}_1(x) \rep{X}{3}_2(x +
      \frac{\im}{2})} \cdot \frac{\varphi_2 \rep{\chi}{2}_2}{\rep{\chi}{1}_1 \rep{\chi}{3}_2}\msp,\\
    \rep{B}{2}_3(x) &= \frac{\rep{X}{1}_1(x) \rep{X}{1}_2(x)}{q_2^{(h)}(x)
      \rep{\widetilde\Lambda}{1}(x)} \cdot \frac{\rep{\chi}{1}_1
      \rep{\chi}{1}_2}{\varphi_2 \rep{a}{1}}\msp,\\
    \rep{B}{2}_4(x) &= \frac{\rep{X}{3}_1(x-\frac{\im}{2})
      \rep{X}{3}_2(x+\frac{\im}{2})}{q_2^{(h)}(x) \rep{\widetilde\Lambda}{3}(x)}
    \cdot \frac{\rep{\chi}{3}_1 \rep{\chi}{3}_2}{\varphi_2 \rep{a}{3}}\msp,\\
    \rep{B}{2}_5(x) &= \frac{q_2^{(h)}(x)
      \rep{X}{2}_1(x-\frac{\im}{2})}{q_2(x-\im) \rep{X}{1}_2(x)
      \rep{X}{3}_1(x-\frac{\im}{2})} \cdot \frac{\varphi_2
      \rep{\chi}{2}_1}{\rep{\chi}{1}_2 \rep{\chi}{3}_1}\msp,\\
    \rep{B}{2}_6(x) &= \frac{q_2(x-\im)
      \rep{\widetilde\Lambda}{2}(x-\frac{\im}{2})}{\rep{X}{2}_1(x-\frac{\im}{2})}
    \cdot \frac{\rep{a}{2}}{\rep{\chi}{2}_1}\msp,\\
    \rep{B}{3}_1(x) &= \frac{q_3(x+\frac{3}{2}\im)
      \rep{\widetilde\Lambda}{3}(x+\frac{\im}{2})}{\phi_+(x+\frac{5}{2}\im)
      \rep{X}{3}_1(x)} \cdot \frac{\rep{a}{3}}{\ex{\beta\mu_4} \rep{\chi}{3}_1}\msp,\\
    \rep{B}{3}_2(x) &= \frac{q_3^{(h)}(x+\frac{\im}{2})
      \rep{X}{3}_1(x)}{q_3(x+\frac{3}{2}\im) \rep{X}{2}_1(x)} \cdot
    \frac{\varphi_3 \rep{\chi}{3}_1}{\ex{\beta\mu_3} \rep{\chi}{2}_1}\msp,\\
    \rep{B}{3}_3(x) &=\frac{\rep{X}{2}_1(x) \rep{X}{3}_2(x)}{q_1(x-\frac{3}{2}\im) q_3^{(h)}(x+\frac{\im}{2})
      \rep{\widetilde\Lambda}{2}(x)} \cdot \frac{\rep{\chi}{2}_1
      \rep{\chi}{3}_2}{\ex{\beta\mu_2} \varphi_3 \rep{a}{2}}\msp,\\
    \rep{B}{3}_4(x) &=\frac{q_1(x-\frac{3}{2}\im)
      \rep{\widetilde\Lambda}{3}_1(x-\frac{\im}{2})}{\phi_-(x-\frac{5}{2}\im)
      \rep{X}{3}_2(x)} \cdot \frac{\rep{a}{3}}{\ex{\beta\mu_1} \rep{\chi}{3}_2}\msp.
  \end{align}
\end{subequations}
Now, we apply~\eqref{eq:logft} to all auxiliary functions. For brevity, we just
treat the case $k < 0$ here as the calculation is completely analogous for $k
> 0$. In this case, we just have to deal with roots and poles, which are
located above the real axis. For the functions $\rep{b}{n}_j(x)$ we get 
the result
\begin{subequations}\label{eq:auxft1}
  \begin{align}
    \rep{\widehat b}{1}_1(k) &= \ex{k/2} \widehat\phi_-(k) - \ex{-k/2}
    \rep{\widehat X}{1}_2(k)\msp,\\
    \rep{\widehat b}{1}_2(k) &= \ex{k/2} \widehat\phi_-(k) +
    \rep{\widehat X}{3}_2(k) - \ex{-k/2} \widehat q_3^{(h)}(k) -
    \rep{\widehat\Lambda}{2}(k)\msp,\\
    \rep{\widehat b}{1}_3(k) &= \ex{k/2} \widehat\phi_-(k) + \ex{3k/2}
    \widehat q_2(k) - \ex{k/2} \widehat q_3(k) - \rep{\widehat
      X}{2}_2(k)\msp,\\
    \rep{\widehat b}{1}_4(k) &= \ex{k/2} \widehat\phi_-(k) + \ex{3k/2}
    \widehat q_3(k) - \ex{k/2} \widehat\phi_+(k) - \ex{k/2}
    \rep{\widehat X}{1}_1(k)\msp,\\
    \rep{\widehat b}{2}_1(k) &= \ex{k} \widehat\phi_-(k) - \ex{-k/2}
    \rep{\widehat X}{2}_2(k)\msp,\\
    \rep{\widehat b}{2}_2(k) &= \ex{k} \widehat\phi_-(k) + \ex{k} \widehat
    q_2(k) - \rep{\widehat X}{1}_1(k) - \ex{-k/2} \rep{\widehat
      X}{3}_2(k)\msp,\\
    \rep{\widehat b}{2}_3(k) &= \ex{k} \widehat\phi_-(k) + \ex{k} \widehat
    q_3(k) - \widehat q_2^{(h)}(k) - \rep{\widehat\Lambda}{1}(k)\msp,\\
    \rep{\widehat b}{2}_4(k) &= \ex{k} \widehat\phi_-(k) + \ex{2k} \widehat
    q_1(k) - \widehat q_2^{(h)}(k) - \rep{\widehat\Lambda}{3}(k)\msp,\\
    \rep{\widehat b}{2}_5(k) &= \ex{k} \widehat\phi_-(k) + \ex{2k} \widehat
    q_1(k) + \ex{k} \widehat q_3(k) - \ex{k} \widehat q_2(k) - \rep{\widehat
      X}{1}_2(k) - \ex{k/2} \rep{\widehat X}{3}_1(k)\msp,\\
    \rep{\widehat b}{2}_6(k) &= \ex{k} \widehat\phi_-(k) + \ex{2 k} \widehat
    q_2(k) - \ex{k/2} \rep{\widehat X}{2}_1(k)\msp,\\
    \rep{\widehat b}{3}_1(k) &= \ex{3k/2} \widehat\phi_-(k) - \rep{\widehat
      X}{3}_1(k)\msp,\\
    \rep{\widehat b}{3}_2(k) &= \ex{3k/2} \widehat\phi_-(k) + \ex{k/2}
    \widehat q_3(k) - \rep{\widehat X}{2}_1(k)\msp,\\
    \begin{split}
      \rep{\widehat b}{3}_3(k) &= \ex{3k/2} \widehat\phi_-(k) + \ex{3k/2}
      \widehat q_2(k) + \ex{-k/2} \rep{\widehat X}{1}_2(k)\\
      & \qquad- \ex{3k/2} \widehat q_1(k) - \ex{-k/2} \widehat q_3^{(h)}(k) -
      \rep{\widehat\Lambda}{2}(k)\msp,
    \end{split}\\
    \rep{\widehat b}{3}_4(k) &= \ex{3k/2} \widehat\phi_-(k) + \ex{5k/2}
    \widehat q_1(k) - \ex{5k/2} \widehat\phi_-(k) - \rep{\widehat X}{3}_2(k)\msp.
  \end{align}
\end{subequations}
For the uppercase functions $\rep{B}{n}_j(x)$ we find
\begin{subequations}\label{eq:auxft2}
  \begin{align}
    \rep{\widehat B}{1}_1(k) &= \ex{-k/2}\rep{\widehat\Lambda}{1}(k) - \ex{-k/2}
    \rep{\widehat X}{1}_2(k)\msp,\\
    \rep{\widehat B}{1}_2(k) &= \ex{-k/2} \rep{\widehat X}{1}_2(k) +
    \rep{\widehat X}{2}_2(k) - \ex{-k/2} \widehat q_3^{(h)}(k) -
    \rep{\widehat\Lambda}{2}(k)\msp,\\
    \rep{\widehat B}{1}_3(k) &= \ex{-k/2} \widehat q_3^{(h)}(k) + \ex{k/2}
    \rep{\widehat X}{1}_1(k) - \ex{k/2} \widehat q_3(k) - \rep{\widehat
      X}{2}_2(k)\msp,\\
    \rep{\widehat B}{1}_4(k) &= \ex{k/2} \widehat q_3(k) + \ex{k/2}
    \rep{\widehat\Lambda}{1}(k) - \ex{k/2} \widehat\phi_+(k) - \ex{k/2}
    \rep{\widehat X}{1}_1(k)\msp,\\
    \rep{\widehat B}{2}_1(k) &= \ex{-k/2} \rep{\widehat\Lambda}{2}(k) - \ex{-k/2}
    \rep{\widehat X}{2}_2(k)\msp,\\
    \rep{\widehat B}{2}_2(k) &= \widehat q_2^{(h)}(k) + \ex{-k/2}
    \rep{\widehat X}{2}_2(k) - \rep{\widehat X}{1}_1(k) - \ex{-k/2} \rep{\widehat
      X}{3}_2(k)\msp,\\
    \rep{\widehat B}{2}_3(k) &= \rep{\widehat X}{1}_1(k) + \rep{\widehat
      X}{1}_2(k) - \widehat q_2^{(h)}(k) - \rep{\widehat\Lambda}{1}(k)\msp,\\
    \rep{\widehat B}{2}_4(k) &= \ex{k/2} \rep{\widehat X}{3}_1(k)  + \ex{-k/2}
    \rep{\widehat X}{3}_2(k) - \widehat q_2^{(h)}(k) - \rep{\widehat\Lambda}{3}(k)\msp,\\
    \rep{\widehat B}{2}_5(k) &= \widehat q_2^{(h)}(k) + \ex{k/2} \rep{\widehat
      X}{2}_1(k) - \ex{k} \widehat q_2(k) - \rep{\widehat
      X}{1}_2(k) - \ex{k/2} \rep{\widehat X}{3}_1(k)\msp,\\
    \rep{\widehat B}{2}_6(k) &= \ex{k} \widehat q_2(k) + \ex{k/2}
    \rep{\widehat\Lambda}{2}(k) - \ex{k/2} \rep{\widehat X}{2}_1(k)\msp,\\
    \rep{\widehat B}{3}_1(k) &= \ex{-k/2} \rep{\widehat\Lambda}{3}(k)  - \rep{\widehat
      X}{3}_1(k)\msp,\\
    \rep{\widehat B}{3}_2(k) &= \ex{-k/2} \widehat q_3^{(h)}(k) +
    \rep{\widehat X}{3}_1(k) - \rep{\widehat X}{2}_1(k)\msp,\\
      \rep{\widehat B}{3}_3(k) &= \rep{\widehat X}{2}_1(k) + \rep{\widehat
        X}{3}_2(k) - \ex{3k/2} \widehat q_1(k) - \ex{-k/2} \widehat q_3^{(h)}(k) -
      \rep{\widehat\Lambda}{2}(k)\msp,\\
    \rep{\widehat B}{3}_4(k) &= \ex{3 k/2} \widehat q_1(k) + \ex{k/2}
    \rep{\widehat\Lambda}{3}(k) - \ex{5k/2} \widehat\phi_-(k) - \rep{\widehat
      X}{3}_2(k)\msp.
  \end{align}
\end{subequations}
The crucial point is the observation that the latter forms a system
of 14 linear equations, exactly as many as there are unknown functions
besides $\rep{\widehat B}{n}_j(k)$. Therefore, we can solve~\eqref{eq:auxft2}
to get $\widehat q_1(k)$, $\widehat q_2(k)$, $\widehat
q_3(k)$, $\rep{\widehat\Lambda}{1}(k)$, $\rep{\widehat\Lambda}{2}(k)$,
$\rep{\widehat\Lambda}{3}(k)$, $\widehat q_2^{(h)}(k)$, $\widehat q_3^{(h)}(k)$, $\rep{\widehat
  X}{1}_1(k)$, $\rep{\widehat X}{1}_2(k)$, $\rep{\widehat X}{2}_1(k)$,
$\rep{\widehat X}{2}_2(k)$, $\rep{\widehat X}{3}_1(k)$ and $\rep{\widehat
  X}{3}_2(k)$ in terms of the auxiliary functions $\rep{\widehat
  B}{n}_j(k)$. Eventually, we substitute this result into~\eqref{eq:auxft1} and
are left with a set of 14 equations, in which only $\rep{\widehat
  b}{n}_j(k)$ and $\rep{\widehat B}{n}_j(k)$ appear.

We combine the results from the cases $k < 0$ and $k > 0$ to get a system of
equations valid for all $k \in \set{R}$. We find the equations to be
\begin{equation}\label{eq:nlieft}
  \vecl{\widehat b}(k) = -\im N \sinh(k \beta/N) \vecl{\widehat V}(k) + \matr{\widehat K}(k) \cdot
  \vecl{\widehat B}(k)\msp,
\end{equation}
where
\begin{align}
  \vecl{\widehat b} &= \left(\rep{\widehat b}{1}_1, \ldots, \rep{\widehat
      b}{1}_4, \rep{\widehat b}{2}_1, \ldots, \rep{\widehat b}{2}_6, \rep{\widehat
      b}{3}_1, \ldots, \rep{\widehat b}{3}_4\right)^\mathrm T\msp,\\
  \vecl{\widehat B} &= \left(\rep{\widehat B}{1}_1, \ldots, \rep{\widehat
      B}{1}_4, \rep{\widehat B}{2}_1, \ldots, \rep{\widehat B}{2}_6,
    \rep{\widehat B}{3}_1, \ldots, \rep{\widehat B}{3}_4\right)^\mathrm
  T\msp,\\
  \vect{\widehat V} &= \left(\rep{\widehat V}{1}_{[4]}, \ldots, \rep{\widehat
      V}{1}_{[4]}, \rep{\widehat V}{2}_{[4]}, \ldots, \rep{\widehat
      V}{2}_{[4]}, \rep{\widehat V}{3}_{[4]}, \ldots, \rep{\widehat
      V}{3}_{[4]}\right)^\mathrm T\msp.
\end{align}
The matrix $\matr{\widehat K}(k)$ is exactly the Fourier transform of the
kernel matrix~\eqref{eq:nliekernel}, and the function
\begin{equation}
  \rep{\widehat V}{j}_{[q]}(k) = \frac{\sinh([q - j] k / 2)}{\sinh(q k/2)}
\end{equation}
is the Fourier transform of~\eqref{eq:nlievdriv}. Only the first term on the
right-hand side of~\eqref{eq:nlieft} contains the Trotter number $N$
explicitly. To analytically perform the global Trotter limit $N \to \infty$,
we therefore just have to consider
\begin{equation}\label{eq:trotterlimit}
  \lim_{N \to \infty} N \sinh(k \beta/N) = k \beta\msp.
\end{equation}
Next, we apply the inverse Fourier transform and
an integration over $x$ to equation~\eqref{eq:nlieft} to eventually obtain the
NLIE~\eqref{eq:nlie}--\eqref{eq:nlievdriv}. The missing integration constants
are determined by considering~\eqref{eq:nlie} in the limit $x \to \infty$. The
asymptotics of the auxiliary functions can be easily read off
from~\eqref{eq:auxfact1} and \eqref{eq:auxfact2}, because only the constant
factors on the right-hand sides survive for large $x$. For the convolutions
with the kernel functions, we find
\begin{equation}
  \lim_{x \to \infty}\bigl[K_i \ast \rep{B}{n}_j\bigr](x) =
  \rep{B}{n}_j(\infty) \int_{-\infty}^{\infty} K_i(x) \,\frac{\dif x}{2 \pi} =
  \rep{B}{n}_j(\infty) \widehat K_i(0)\msp.
\end{equation}
Inserting this information into~\eqref{eq:nlie} finally leads to the constants
given in~\eqref{eq:nlieconstants}.

To also derive the formula for the largest eigenvalue~\eqref{eq:nlielambda},
one has to recall that the eigenvalue already appeared during the
calculation. Its logarithmic Fourier transform $\rep{\widehat\Lambda}{1}(k)$
could be expressed solely in terms of $\rep{\widehat B}{n}_j(k)$. We
additionally define the function
\begin{equation}
  \rep{\underline\Lambda}{1}(x) = \frac{\rep{\Lambda}{1}(x)}{\phi_-(x-\im)
    \phi_+(x+\im)}\msp,
\end{equation}
which has the advantage of having constant asymptotics. In
the Trotter limit $N \to \infty$, this definition simply yields $\ln\rep{\Lambda}{1}(0) =
\ln\rep{\underline\Lambda}{1}(0) - \beta$. Using the previously obtained information on
$\rep{\widehat\Lambda}{1}(k)$, we can write
\begin{equation}
  \rep{\widehat{\underline\Lambda}}{1}(k) = \im N \sinh(k \beta/N)
  \ex{-|k|/2} \frac{\sinh(3 k/2)}{\sinh(2 k)} + \sum_{n=1}^3
  \sum_{j=1}^{d_n} \rep{\widehat V}{n}_{[4]}(k) \cdot \ln\rep{\widehat B}{n}_j(k)\msp.
\end{equation}
Now we proceed as above. To perform the Trotter limit, we just have to
use~\eqref{eq:trotterlimit}. Then we apply the inverse Fourier
transform and an integration over the spectral parameter. Again, we find the
integration constant by considering the limit $x \to \infty$. We finally arrive
at equation~\eqref{eq:nlielambda}.

\phantomsection
\addcontentsline{toc}{section}{\refname}
\bibliography{su4paper}

\providecommand{\alg}[1]{\mathit{#1}}
\begin{thebibliography}{60}
\expandafter\ifx\csname natexlab\endcsname\relax\def\natexlab#1{#1}\fi
\expandafter\ifx\csname bibnamefont\endcsname\relax
  \def\bibnamefont#1{#1}\fi
\expandafter\ifx\csname bibfnamefont\endcsname\relax
  \def\bibfnamefont#1{#1}\fi
\expandafter\ifx\csname citenamefont\endcsname\relax
  \def\citenamefont#1{#1}\fi
\expandafter\ifx\csname url\endcsname\relax
  \def\url#1{\texttt{#1}}\fi
\expandafter\ifx\csname urlprefix\endcsname\relax\def\urlprefix{URL }\fi
\providecommand{\bibinfo}[2]{#2}
\providecommand{\eprint}[2][]{\url{#2}}

\bibitem[{\citenamefont{Bethe}(1931)}]{bethe31}
\bibinfo{author}{\bibfnamefont{H.}~\bibnamefont{Bethe}}, \bibinfo{journal}{Z.
  Phys.} \textbf{\bibinfo{volume}{71}}, \bibinfo{pages}{205}
  (\bibinfo{year}{1931}).

\bibitem[{\citenamefont{Uimin}(1970)}]{uimin70}
\bibinfo{author}{\bibfnamefont{G.~V.} \bibnamefont{Uimin}},
  \bibinfo{journal}{JETP Lett.} \textbf{\bibinfo{volume}{12}},
  \bibinfo{pages}{225} (\bibinfo{year}{1970}).

\bibitem[{\citenamefont{Sutherland}(1975)}]{sutherland75}
\bibinfo{author}{\bibfnamefont{B.}~\bibnamefont{Sutherland}},
  \bibinfo{journal}{Phys. Rev. B} \textbf{\bibinfo{volume}{12}},
  \bibinfo{pages}{3795} (\bibinfo{year}{1975}).

\bibitem[{\citenamefont{Perk and Schultz}(1981)}]{perk81}
\bibinfo{author}{\bibfnamefont{J.~H.~H.} \bibnamefont{Perk}} \bibnamefont{and}
  \bibinfo{author}{\bibfnamefont{C.~L.} \bibnamefont{Schultz}},
  \bibinfo{journal}{Phys. Lett. A} \textbf{\bibinfo{volume}{84}},
  \bibinfo{pages}{407} (\bibinfo{year}{1981}).

\bibitem[{\citenamefont{Andrei and Johannesson}(1984)}]{andrei84}
\bibinfo{author}{\bibfnamefont{N.}~\bibnamefont{Andrei}} \bibnamefont{and}
  \bibinfo{author}{\bibfnamefont{H.}~\bibnamefont{Johannesson}},
  \bibinfo{journal}{Phys. Lett. A} \textbf{\bibinfo{volume}{104}},
  \bibinfo{pages}{370} (\bibinfo{year}{1984}).

\bibitem[{\citenamefont{Johannesson}(1986{\natexlab{a}})}]{johannesson86a}
\bibinfo{author}{\bibfnamefont{H.}~\bibnamefont{Johannesson}},
  \bibinfo{journal}{Nucl. Phys. B} \textbf{\bibinfo{volume}{270}},
  \bibinfo{pages}{235} (\bibinfo{year}{1986}{\natexlab{a}}).

\bibitem[{\citenamefont{Affleck}(1986)}]{affleck86a}
\bibinfo{author}{\bibfnamefont{I.}~\bibnamefont{Affleck}},
  \bibinfo{journal}{Nucl. Phys. B} \textbf{\bibinfo{volume}{265}},
  \bibinfo{pages}{409} (\bibinfo{year}{1986}).

\bibitem[{\citenamefont{Affleck}(1988)}]{affleck88}
\bibinfo{author}{\bibfnamefont{I.}~\bibnamefont{Affleck}},
  \bibinfo{journal}{Nucl. Phys. B} \textbf{\bibinfo{volume}{305}},
  \bibinfo{pages}{582} (\bibinfo{year}{1988}).

\bibitem[{\citenamefont{Yang and Yang}(1969)}]{yang69}
\bibinfo{author}{\bibfnamefont{C.~N.} \bibnamefont{Yang}} \bibnamefont{and}
  \bibinfo{author}{\bibfnamefont{C.~P.} \bibnamefont{Yang}},
  \bibinfo{journal}{J. Math. Phys.} \textbf{\bibinfo{volume}{10}},
  \bibinfo{pages}{1115} (\bibinfo{year}{1969}).

\bibitem[{\citenamefont{Yang}(1970)}]{yang70}
\bibinfo{author}{\bibfnamefont{C.~P.} \bibnamefont{Yang}},
  \bibinfo{journal}{Phys. Rev. A} \textbf{\bibinfo{volume}{2}},
  \bibinfo{pages}{154} (\bibinfo{year}{1970}).

\bibitem[{\citenamefont{Takahashi}(1971)}]{takahashi71}
\bibinfo{author}{\bibfnamefont{M.}~\bibnamefont{Takahashi}},
  \bibinfo{journal}{Prog. Theor. Phys.} \textbf{\bibinfo{volume}{46}},
  \bibinfo{pages}{401} (\bibinfo{year}{1971}).

\bibitem[{\citenamefont{Gaudin}(1971)}]{gaudin71}
\bibinfo{author}{\bibfnamefont{M.}~\bibnamefont{Gaudin}},
  \bibinfo{journal}{Phys. Rev. Lett.} \textbf{\bibinfo{volume}{26}},
  \bibinfo{pages}{1301} (\bibinfo{year}{1971}).

\bibitem[{\citenamefont{Johannesson}(1986{\natexlab{b}})}]{johannesson86b}
\bibinfo{author}{\bibfnamefont{H.}~\bibnamefont{Johannesson}},
  \bibinfo{journal}{Phys. Lett. A} \textbf{\bibinfo{volume}{116}},
  \bibinfo{pages}{133} (\bibinfo{year}{1986}{\natexlab{b}}).

\bibitem[{\citenamefont{Schlottmann}(1992)}]{schlottmann92}
\bibinfo{author}{\bibfnamefont{P.}~\bibnamefont{Schlottmann}},
  \bibinfo{journal}{Phys. Rev. B} \textbf{\bibinfo{volume}{45}},
  \bibinfo{pages}{5293} (\bibinfo{year}{1992}).

\bibitem[{\citenamefont{Lee}(1994{\natexlab{a}})}]{lee94a}
\bibinfo{author}{\bibfnamefont{K.}~\bibnamefont{Lee}}, \bibinfo{journal}{J.
  Korean Phys. Soc.} \textbf{\bibinfo{volume}{27}}, \bibinfo{pages}{205}
  (\bibinfo{year}{1994}{\natexlab{a}}).

\bibitem[{\citenamefont{Lee}(1994{\natexlab{b}})}]{lee94b}
\bibinfo{author}{\bibfnamefont{K.}~\bibnamefont{Lee}}, \bibinfo{journal}{Phys.
  Lett. A} \textbf{\bibinfo{volume}{187}}, \bibinfo{pages}{112}
  (\bibinfo{year}{1994}{\natexlab{b}}).

\bibitem[{\citenamefont{Suzuki}(1985)}]{suzuki85}
\bibinfo{author}{\bibfnamefont{M.}~\bibnamefont{Suzuki}},
  \bibinfo{journal}{Phys. Rev. B} \textbf{\bibinfo{volume}{31}},
  \bibinfo{pages}{2957} (\bibinfo{year}{1985}).

\bibitem[{\citenamefont{Kl{\"u}mper}(1992)}]{kluemper92b}
\bibinfo{author}{\bibfnamefont{A.}~\bibnamefont{Kl{\"u}mper}},
  \bibinfo{journal}{Ann. Phys. (Leipzig)} \textbf{\bibinfo{volume}{1}},
  \bibinfo{pages}{540} (\bibinfo{year}{1992}).

\bibitem[{\citenamefont{Kirillov and Reshetikhin}(1987)}]{kirillov87}
\bibinfo{author}{\bibfnamefont{A.~N.} \bibnamefont{Kirillov}} \bibnamefont{and}
  \bibinfo{author}{\bibfnamefont{N.~Y.} \bibnamefont{Reshetikhin}},
  \bibinfo{journal}{J. Phys. A: Math. Gen.} \textbf{\bibinfo{volume}{20}},
  \bibinfo{pages}{1565} (\bibinfo{year}{1987}).

\bibitem[{\citenamefont{Bazhanov and Reshetikhin}(1990)}]{bazhanov90}
\bibinfo{author}{\bibfnamefont{V.~V.} \bibnamefont{Bazhanov}} \bibnamefont{and}
  \bibinfo{author}{\bibfnamefont{N.~Y.} \bibnamefont{Reshetikhin}},
  \bibinfo{journal}{J. Phys. A: Math. Gen.} \textbf{\bibinfo{volume}{23}},
  \bibinfo{pages}{1477} (\bibinfo{year}{1990}).

\bibitem[{\citenamefont{Kl{\"u}mper and Pearce}(1992)}]{kluemper92a}
\bibinfo{author}{\bibfnamefont{A.}~\bibnamefont{Kl{\"u}mper}} \bibnamefont{and}
  \bibinfo{author}{\bibfnamefont{P.~A.} \bibnamefont{Pearce}},
  \bibinfo{journal}{Physica A} \textbf{\bibinfo{volume}{183}},
  \bibinfo{pages}{304} (\bibinfo{year}{1992}).

\bibitem[{\citenamefont{Kuniba et~al.}(1994)\citenamefont{Kuniba, Nakanishi,
  and Suzuki}}]{kuniba94}
\bibinfo{author}{\bibfnamefont{A.}~\bibnamefont{Kuniba}},
  \bibinfo{author}{\bibfnamefont{T.}~\bibnamefont{Nakanishi}},
  \bibnamefont{and} \bibinfo{author}{\bibfnamefont{J.}~\bibnamefont{Suzuki}},
  \bibinfo{journal}{Int. J. Mod. Phys. A} \textbf{\bibinfo{volume}{9}},
  \bibinfo{pages}{5215} (\bibinfo{year}{1994}).

\bibitem[{\citenamefont{Tsuboi}(1997)}]{tsuboi97}
\bibinfo{author}{\bibfnamefont{Z.}~\bibnamefont{Tsuboi}}, \bibinfo{journal}{J.
  Phys. A: Math. Gen.} \textbf{\bibinfo{volume}{30}}, \bibinfo{pages}{7975}
  (\bibinfo{year}{1997}).

\bibitem[{\citenamefont{Tsuboi}(1998)}]{tsuboi98}
\bibinfo{author}{\bibfnamefont{Z.}~\bibnamefont{Tsuboi}},
  \bibinfo{journal}{Physica A} \textbf{\bibinfo{volume}{252}},
  \bibinfo{pages}{565} (\bibinfo{year}{1998}).

\bibitem[{\citenamefont{J{\"u}ttner et~al.}(1998)\citenamefont{J{\"u}ttner,
  Kl{\"u}mper, and Suzuki}}]{juettner98}
\bibinfo{author}{\bibfnamefont{G.}~\bibnamefont{J{\"u}ttner}},
  \bibinfo{author}{\bibfnamefont{A.}~\bibnamefont{Kl{\"u}mper}},
  \bibnamefont{and} \bibinfo{author}{\bibfnamefont{J.}~\bibnamefont{Suzuki}},
  \bibinfo{journal}{Nucl. Phys. B} \textbf{\bibinfo{volume}{512}},
  \bibinfo{pages}{581} (\bibinfo{year}{1998}).

\bibitem[{\citenamefont{Kl{\"u}mper and Batchelor}(1990)}]{kluemper90}
\bibinfo{author}{\bibfnamefont{A.}~\bibnamefont{Kl{\"u}mper}} \bibnamefont{and}
  \bibinfo{author}{\bibfnamefont{M.~T.} \bibnamefont{Batchelor}},
  \bibinfo{journal}{J. Phys. A: Math. Gen.} \textbf{\bibinfo{volume}{23}},
  \bibinfo{pages}{L189} (\bibinfo{year}{1990}).

\bibitem[{\citenamefont{Kl{\"u}mper et~al.}(1991)\citenamefont{Kl{\"u}mper,
  Batchelor, and Pearce}}]{kluemper91}
\bibinfo{author}{\bibfnamefont{A.}~\bibnamefont{Kl{\"u}mper}},
  \bibinfo{author}{\bibfnamefont{M.~T.} \bibnamefont{Batchelor}},
  \bibnamefont{and} \bibinfo{author}{\bibfnamefont{P.~A.}
  \bibnamefont{Pearce}}, \bibinfo{journal}{J. Phys. A: Math. Gen.}
  \textbf{\bibinfo{volume}{24}}, \bibinfo{pages}{3111} (\bibinfo{year}{1991}).

\bibitem[{\citenamefont{Destri and de~Vega}(1992)}]{destri92}
\bibinfo{author}{\bibfnamefont{C.}~\bibnamefont{Destri}} \bibnamefont{and}
  \bibinfo{author}{\bibfnamefont{H.~J.} \bibnamefont{de~Vega}},
  \bibinfo{journal}{Phys. Rev. Lett.} \textbf{\bibinfo{volume}{69}},
  \bibinfo{pages}{2313} (\bibinfo{year}{1992}).

\bibitem[{\citenamefont{Kl{\"u}mper}(1993)}]{kluemper93}
\bibinfo{author}{\bibfnamefont{A.}~\bibnamefont{Kl{\"u}mper}},
  \bibinfo{journal}{Z. Phys. B} \textbf{\bibinfo{volume}{91}},
  \bibinfo{pages}{507} (\bibinfo{year}{1993}).

\bibitem[{\citenamefont{Suzuki}(1999)}]{suzuki99}
\bibinfo{author}{\bibfnamefont{J.}~\bibnamefont{Suzuki}}, \bibinfo{journal}{J.
  Phys. A: Math. Gen.} \textbf{\bibinfo{volume}{32}}, \bibinfo{pages}{2341}
  (\bibinfo{year}{1999}).

\bibitem[{\citenamefont{J{\"u}ttner and Kl{\"u}mper}(1997)}]{juettner97a}
\bibinfo{author}{\bibfnamefont{G.}~\bibnamefont{J{\"u}ttner}} \bibnamefont{and}
  \bibinfo{author}{\bibfnamefont{A.}~\bibnamefont{Kl{\"u}mper}},
  \bibinfo{journal}{Europhys. Lett.} \textbf{\bibinfo{volume}{37}},
  \bibinfo{pages}{335} (\bibinfo{year}{1997}).

\bibitem[{\citenamefont{J{\"u}ttner et~al.}(1997)\citenamefont{J{\"u}ttner,
  Kl{\"u}mper, and Suzuki}}]{juettner97b}
\bibinfo{author}{\bibfnamefont{G.}~\bibnamefont{J{\"u}ttner}},
  \bibinfo{author}{\bibfnamefont{A.}~\bibnamefont{Kl{\"u}mper}},
  \bibnamefont{and} \bibinfo{author}{\bibfnamefont{J.}~\bibnamefont{Suzuki}},
  \bibinfo{journal}{Nucl. Phys. B} \textbf{\bibinfo{volume}{487}},
  \bibinfo{pages}{650} (\bibinfo{year}{1997}).

\bibitem[{\citenamefont{Fujii and Kl{\"u}mper}(1999)}]{fujii99}
\bibinfo{author}{\bibfnamefont{A.}~\bibnamefont{Fujii}} \bibnamefont{and}
  \bibinfo{author}{\bibfnamefont{A.}~\bibnamefont{Kl{\"u}mper}},
  \bibinfo{journal}{Nucl. Phys. B} \textbf{\bibinfo{volume}{546}},
  \bibinfo{pages}{751} (\bibinfo{year}{1999}).

\bibitem[{\citenamefont{Takahashi}(2001)}]{takahashi01}
\bibinfo{author}{\bibfnamefont{M.}~\bibnamefont{Takahashi}}, in
  \emph{\bibinfo{booktitle}{Physics and Combinatorics 2000}}, edited by
  \bibinfo{editor}{\bibfnamefont{A.~N.} \bibnamefont{Kirillov}}
  \bibnamefont{and} \bibinfo{editor}{\bibfnamefont{N.}~\bibnamefont{Liskova}}
  (\bibinfo{publisher}{World Scientific}, \bibinfo{address}{Singapore},
  \bibinfo{year}{2001}), pp. \bibinfo{pages}{299--304}.

\bibitem[{\citenamefont{Tsuboi}(2003)}]{tsuboi03}
\bibinfo{author}{\bibfnamefont{Z.}~\bibnamefont{Tsuboi}}, \bibinfo{journal}{J.
  Phys. A: Math. Gen.} \textbf{\bibinfo{volume}{36}}, \bibinfo{pages}{1493}
  (\bibinfo{year}{2003}).

\bibitem[{\citenamefont{Shiroishi and Takahashi}(2002)}]{shiroishi02}
\bibinfo{author}{\bibfnamefont{M.}~\bibnamefont{Shiroishi}} \bibnamefont{and}
  \bibinfo{author}{\bibfnamefont{M.}~\bibnamefont{Takahashi}},
  \bibinfo{journal}{Phys. Rev. Lett.} \textbf{\bibinfo{volume}{89}},
  \bibinfo{pages}{117201} (\bibinfo{year}{2002}).

\bibitem[{\citenamefont{Tsuboi}(2006)}]{tsuboi06}
\bibinfo{author}{\bibfnamefont{Z.}~\bibnamefont{Tsuboi}},
  \bibinfo{journal}{Nucl. Phys. B} \textbf{\bibinfo{volume}{737}},
  \bibinfo{pages}{261} (\bibinfo{year}{2006}).

\bibitem[{\citenamefont{Yamashita et~al.}(1998)\citenamefont{Yamashita,
  Shibata, and Ueda}}]{yamashita98}
\bibinfo{author}{\bibfnamefont{Y.}~\bibnamefont{Yamashita}},
  \bibinfo{author}{\bibfnamefont{N.}~\bibnamefont{Shibata}}, \bibnamefont{and}
  \bibinfo{author}{\bibfnamefont{K.}~\bibnamefont{Ueda}},
  \bibinfo{journal}{Phys. Rev. B} \textbf{\bibinfo{volume}{58}},
  \bibinfo{pages}{9114} (\bibinfo{year}{1998}).

\bibitem[{\citenamefont{Yamashita et~al.}(2000)\citenamefont{Yamashita,
  Shibata, and Ueda}}]{yamashita00}
\bibinfo{author}{\bibfnamefont{Y.}~\bibnamefont{Yamashita}},
  \bibinfo{author}{\bibfnamefont{N.}~\bibnamefont{Shibata}}, \bibnamefont{and}
  \bibinfo{author}{\bibfnamefont{K.}~\bibnamefont{Ueda}},
  \bibinfo{journal}{Phys. Rev. B} \textbf{\bibinfo{volume}{61}},
  \bibinfo{pages}{4012} (\bibinfo{year}{2000}).

\bibitem[{\citenamefont{Frischmuth et~al.}(1999)\citenamefont{Frischmuth, Mila,
  and Troyer}}]{frischmuth99}
\bibinfo{author}{\bibfnamefont{B.}~\bibnamefont{Frischmuth}},
  \bibinfo{author}{\bibfnamefont{F.}~\bibnamefont{Mila}}, \bibnamefont{and}
  \bibinfo{author}{\bibfnamefont{M.}~\bibnamefont{Troyer}},
  \bibinfo{journal}{Phys. Rev. Lett.} \textbf{\bibinfo{volume}{82}},
  \bibinfo{pages}{835} (\bibinfo{year}{1999}).

\bibitem[{\citenamefont{Fukushima}(2002)}]{fukushima02}
\bibinfo{author}{\bibfnamefont{N.}~\bibnamefont{Fukushima}},
  \bibinfo{journal}{J. Phys. Soc. Jpn.} \textbf{\bibinfo{volume}{71}},
  \bibinfo{pages}{1238} (\bibinfo{year}{2002}).

\bibitem[{\citenamefont{Sirker}(2004)}]{sirker04}
\bibinfo{author}{\bibfnamefont{J.}~\bibnamefont{Sirker}},
  \bibinfo{journal}{Phys. Rev. B} \textbf{\bibinfo{volume}{69}},
  \bibinfo{pages}{104428} (\bibinfo{year}{2004}).

\bibitem[{\citenamefont{Gu and Li}(2002)}]{gu02}
\bibinfo{author}{\bibfnamefont{S.-J.} \bibnamefont{Gu}} \bibnamefont{and}
  \bibinfo{author}{\bibfnamefont{Y.-Q.} \bibnamefont{Li}},
  \bibinfo{journal}{Phys. Rev. B} \textbf{\bibinfo{volume}{66}},
  \bibinfo{pages}{092404} (\bibinfo{year}{2002}).

\bibitem[{\citenamefont{Wang}(1999)}]{wang99}
\bibinfo{author}{\bibfnamefont{Y.}~\bibnamefont{Wang}}, \bibinfo{journal}{Phys.
  Rev. B} \textbf{\bibinfo{volume}{60}}, \bibinfo{pages}{9236}
  (\bibinfo{year}{1999}).

\bibitem[{\citenamefont{Batchelor
  et~al.}(2003{\natexlab{a}})\citenamefont{Batchelor, Guan, Foerster, and
  Zhou}}]{batchelor03a}
\bibinfo{author}{\bibfnamefont{M.~T.} \bibnamefont{Batchelor}},
  \bibinfo{author}{\bibfnamefont{X.-W.} \bibnamefont{Guan}},
  \bibinfo{author}{\bibfnamefont{A.}~\bibnamefont{Foerster}}, \bibnamefont{and}
  \bibinfo{author}{\bibfnamefont{H.-Q.} \bibnamefont{Zhou}},
  \bibinfo{journal}{New J. Phys.} \textbf{\bibinfo{volume}{5}},
  \bibinfo{pages}{107} (\bibinfo{year}{2003}{\natexlab{a}}).

\bibitem[{\citenamefont{Batchelor
  et~al.}(2003{\natexlab{b}})\citenamefont{Batchelor, Guan, Oelkers, Sakai,
  Tsuboi, and Foerster}}]{batchelor03b}
\bibinfo{author}{\bibfnamefont{M.~T.} \bibnamefont{Batchelor}},
  \bibinfo{author}{\bibfnamefont{X.-W.} \bibnamefont{Guan}},
  \bibinfo{author}{\bibfnamefont{N.}~\bibnamefont{Oelkers}},
  \bibinfo{author}{\bibfnamefont{K.}~\bibnamefont{Sakai}},
  \bibinfo{author}{\bibfnamefont{Z.}~\bibnamefont{Tsuboi}}, \bibnamefont{and}
  \bibinfo{author}{\bibfnamefont{A.}~\bibnamefont{Foerster}},
  \bibinfo{journal}{Phys. Rev. Lett.} \textbf{\bibinfo{volume}{91}},
  \bibinfo{pages}{217202} (\bibinfo{year}{2003}{\natexlab{b}}).

\bibitem[{\citenamefont{Baxter}(1982)}]{baxter82}
\bibinfo{author}{\bibfnamefont{R.~J.} \bibnamefont{Baxter}},
  \emph{\bibinfo{title}{Exactly Solved Models in Statistical Mechanics}}
  (\bibinfo{publisher}{Academic Press}, \bibinfo{address}{London},
  \bibinfo{year}{1982}).

\bibitem[{\citenamefont{Suzuki and Inoue}(1987)}]{suzuki87}
\bibinfo{author}{\bibfnamefont{M.}~\bibnamefont{Suzuki}} \bibnamefont{and}
  \bibinfo{author}{\bibfnamefont{M.}~\bibnamefont{Inoue}},
  \bibinfo{journal}{Prog. Theor. Phys.} \textbf{\bibinfo{volume}{78}},
  \bibinfo{pages}{787} (\bibinfo{year}{1987}).

\bibitem[{\citenamefont{Suzuki et~al.}(1990)\citenamefont{Suzuki, Akutsu, and
  Wadati}}]{suzuki90}
\bibinfo{author}{\bibfnamefont{J.}~\bibnamefont{Suzuki}},
  \bibinfo{author}{\bibfnamefont{Y.}~\bibnamefont{Akutsu}}, \bibnamefont{and}
  \bibinfo{author}{\bibfnamefont{M.}~\bibnamefont{Wadati}},
  \bibinfo{journal}{J. Phys. Soc. Jpn.} \textbf{\bibinfo{volume}{59}},
  \bibinfo{pages}{2667} (\bibinfo{year}{1990}).

\bibitem[{\citenamefont{Kl{\"u}mper et~al.}(1997)\citenamefont{Kl{\"u}mper,
  Wehner, and Zittartz}}]{kluemper97}
\bibinfo{author}{\bibfnamefont{A.}~\bibnamefont{Kl{\"u}mper}},
  \bibinfo{author}{\bibfnamefont{T.}~\bibnamefont{Wehner}}, \bibnamefont{and}
  \bibinfo{author}{\bibfnamefont{J.}~\bibnamefont{Zittartz}},
  \bibinfo{journal}{J. Phys. A: Math. Gen.} \textbf{\bibinfo{volume}{30}},
  \bibinfo{pages}{1897} (\bibinfo{year}{1997}).

\bibitem[{\citenamefont{Suzuki}(1994)}]{suzuki94}
\bibinfo{author}{\bibfnamefont{J.}~\bibnamefont{Suzuki}},
  \bibinfo{journal}{Phys. Lett. A} \textbf{\bibinfo{volume}{195}},
  \bibinfo{pages}{190} (\bibinfo{year}{1994}).

\bibitem[{\citenamefont{Kuniba and Suzuki}(1995)}]{kuniba95a}
\bibinfo{author}{\bibfnamefont{A.}~\bibnamefont{Kuniba}} \bibnamefont{and}
  \bibinfo{author}{\bibfnamefont{J.}~\bibnamefont{Suzuki}},
  \bibinfo{journal}{Commun. Math. Phys.} \textbf{\bibinfo{volume}{173}},
  \bibinfo{pages}{225} (\bibinfo{year}{1995}).

\bibitem[{\citenamefont{Kuniba et~al.}(1995)\citenamefont{Kuniba, Ohta, and
  Suzuki}}]{kuniba95b}
\bibinfo{author}{\bibfnamefont{A.}~\bibnamefont{Kuniba}},
  \bibinfo{author}{\bibfnamefont{Y.}~\bibnamefont{Ohta}}, \bibnamefont{and}
  \bibinfo{author}{\bibfnamefont{J.}~\bibnamefont{Suzuki}},
  \bibinfo{journal}{J. Phys. A: Math. Gen.} \textbf{\bibinfo{volume}{28}},
  \bibinfo{pages}{6211} (\bibinfo{year}{1995}).

\bibitem[{\citenamefont{Kulish and Reshetikhin}(1981)}]{kulish81}
\bibinfo{author}{\bibfnamefont{P.~P.} \bibnamefont{Kulish}} \bibnamefont{and}
  \bibinfo{author}{\bibfnamefont{N.~Y.} \bibnamefont{Reshetikhin}},
  \bibinfo{journal}{Sov. Phys. JETP} \textbf{\bibinfo{volume}{53}},
  \bibinfo{pages}{108} (\bibinfo{year}{1981}).

\bibitem[{\citenamefont{Eggert et~al.}(1994)\citenamefont{Eggert, Affleck, and
  Takahashi}}]{eggert94}
\bibinfo{author}{\bibfnamefont{S.}~\bibnamefont{Eggert}},
  \bibinfo{author}{\bibfnamefont{I.}~\bibnamefont{Affleck}}, \bibnamefont{and}
  \bibinfo{author}{\bibfnamefont{M.}~\bibnamefont{Takahashi}},
  \bibinfo{journal}{Phys. Rev. Lett.} \textbf{\bibinfo{volume}{73}},
  \bibinfo{pages}{332} (\bibinfo{year}{1994}).

\bibitem[{\citenamefont{Kl{\"u}mper}(1998)}]{kluemper98}
\bibinfo{author}{\bibfnamefont{A.}~\bibnamefont{Kl{\"u}mper}},
  \bibinfo{journal}{Euro. Phys. J. B} \textbf{\bibinfo{volume}{5}},
  \bibinfo{pages}{677} (\bibinfo{year}{1998}).

\bibitem[{\citenamefont{Lukyanov}(1998)}]{lukyanov98}
\bibinfo{author}{\bibfnamefont{S.}~\bibnamefont{Lukyanov}},
  \bibinfo{journal}{Nucl. Phys. B} \textbf{\bibinfo{volume}{522}},
  \bibinfo{pages}{533} (\bibinfo{year}{1998}).

\bibitem[{\citenamefont{Kl{\"u}mper and Johnston}(2000)}]{kluemper00}
\bibinfo{author}{\bibfnamefont{A.}~\bibnamefont{Kl{\"u}mper}} \bibnamefont{and}
  \bibinfo{author}{\bibfnamefont{D.~C.} \bibnamefont{Johnston}},
  \bibinfo{journal}{Phys. Rev. Lett.} \textbf{\bibinfo{volume}{84}},
  \bibinfo{pages}{4701} (\bibinfo{year}{2000}).

\bibitem[{\citenamefont{Doikou and Nepomechie}(1998)}]{doikou98}
\bibinfo{author}{\bibfnamefont{A.}~\bibnamefont{Doikou}} \bibnamefont{and}
  \bibinfo{author}{\bibfnamefont{R.~I.} \bibnamefont{Nepomechie}},
  \bibinfo{journal}{Nucl. Phys. B} \textbf{\bibinfo{volume}{521}},
  \bibinfo{pages}{547} (\bibinfo{year}{1998}).

\bibitem[{\citenamefont{Essler et~al.}(1992)\citenamefont{Essler, Korepin, and
  Schoutens}}]{essler92a}
\bibinfo{author}{\bibfnamefont{F.~H.~L.} \bibnamefont{Essler}},
  \bibinfo{author}{\bibfnamefont{V.~E.} \bibnamefont{Korepin}},
  \bibnamefont{and}
  \bibinfo{author}{\bibfnamefont{K.}~\bibnamefont{Schoutens}},
  \bibinfo{journal}{Phys. Rev. Lett.} \textbf{\bibinfo{volume}{68}},
  \bibinfo{pages}{2960} (\bibinfo{year}{1992}).

\end{thebibliography}

\end{document}